\documentclass[twocolumn]{aastex631}
\pdfoutput=1 
\usepackage{amsmath,amstext,amssymb,mathtools,graphicx,color}

\usepackage{graphicx}	
\usepackage{amsmath}	
\usepackage{amssymb}	
\usepackage{xcolor}
\usepackage{lipsum}
\usepackage[normalem]{ulem}
\usepackage{multirow}

\submitjournal{ApJ}

\shorttitle{SBO in Stellar Explosions}
\shortauthors{Fryer et al.}

\graphicspath{{./}{figures/}}

\begin{document}

\title{High-Energy Shock Breakout from Supernovae and Gamma-ray Bursts}

\correspondingauthor{Chris Fryer}
\email{fryer@lanl.gov}

\author[0000-0003-2624-0056]{Christopher~L.~Fryer}
\affiliation{Center for Nonlinear Studies, Los Alamos National Laboratory, Los Alamos, NM 87545 USA}

\author[0000-0002-2942-3379]{Eric~Burns}
\affiliation{Department of Physics \& Astronomy, Louisiana State University, Baton Rouge, LA 70803, USA}

\author[0000-0003-0485-6554]{Joseph~M.~Colosimo}
\affiliation{X-ray Astrophysics Laboratory, NASA Goddard Space Flight Center, Greenbelt, MD 20771, USA}

\author[0000-0002-2942-3379]{Michela~Negro}
\affiliation{Department of Physics \& Astronomy, Louisiana State University, Baton Rouge, LA 70803, USA}

\author[0000-0002-9700-0036]{Brendan O'Connor}
    \altaffiliation{McWilliams Fellow}
    \affiliation{McWilliams Center for Cosmology and Astrophysics, Department of Physics, Carnegie Mellon University, Pittsburgh, PA 15213, USA}

\begin{abstract}

Cosmic explosions play a critical role in a broad range of astrophysical fields. Although considerable progress has been made to understand the explosive engines and their progenitors, many of the details are not well understood. One of the most powerful electromagnetic probes of the explosive mechanism and the stellar progenitor is the first burst of photons emitted from this blastwave as it exits the stellar photosphere, known as shock breakout (SBO). Our understanding of SBO has evolved considerably in the past decade. Shock heating as the blastwave propagates through the star and circumstellar material can drastically alter this emission producing a much broader range of potential SBO signals than that predicted by standard analytical approaches. Here we present a semi-analytic approach to model this diverse SBO emission, focused on thermal Bremsstrahlung radiation, which more accurately captures the complexities in Nature over previous treatments. We calculate a range of signals for a range of supernova and gamma-ray burst types. Our models demonstrate how we can use these signals to place constraints on the nature of the explosive engines and better understand the role SBO can play in prompt gamma-ray bursts. We study the implications of these results to historic observations, Einstein Probe transients, and in the context of proposed missions. We find that stripped envelope events can be detected serendipitously with survey telescopes, but type Ia and II SBO detections require fast-pointing X-ray observations in response to early warning alerts from gravitational wave or neutrino detectors.

\end{abstract}


\section{Introduction} 
\label{sec:intro}

Shock breakout (SBO), or shock emergence, can refer to a broad set of phenomena contributing to pre-peak emission in supernovae (SN) and gamma-ray bursts (GRBs).  In thermonuclear (Type Ia) supernovae (SN), this occurs when the exploding white dwarf (powered by nuclear fusion) expands and becomes optically thin.  For core-collapse supernovae (driven either by neutrinos or jets), SBO occurs when the supernova shock breaks out of the progenitor star's extended stellar envelope (type II SN) or dense stellar wind (Type Ib/c SN).  SBO also occurs in jet-driven explosions (Broad-line Ic SN, long-duration GRBs) when or after the jet breaks out of its massive star.

This field has progressed dramatically in the past 10 years, where both increasingly complex analytic models~\citep{2001ApJ...551..946T,2007ApJ...664.1026W,2010ApJ...708..598P,2010ApJ...719..881S,2017hsn..book..967W,2021MNRAS.508.5766I,2024arXiv241206733I} and hydrodynamic and radiation-hydrodynamics calculations~\citep{2013ApJS..204...16F,2015ApJ...805...98B,2017ApJ...845..103L,2020MNRAS.494.3927K,2020ApJ...898..123F,2022ApJ...933..164G} have demonstrated the complexity of this phenomena.   The original definition of SBO referred to the burst of emission that occurs when an explosive shock reaches the edge of a star in a SN or GRB.  In this definition, the energy arises from the conversion of kinetic energy to thermal energy in the forward shock of a supernova blast wave.   The basic idea of this model can be simply stated.  Although photons carry much of the shock energy, when the shock is in the star, the material is so dense that the photons are trapped in the star, moving outward with the flow of the transient (e.g. supernova) blastwave.  But as the shock reaches the edge of the star, the radiation decouples from the matter, ultimately escaping in a burst of high-energy (UV/X-ray) photons.  

SBO occurs when the radiative velocity (transport velocity, e.g. diffusion velocity in simple models) of the photons exceeds that of the matter velocity of the shock.  The photon spectrum is determined by a blackbody with a temperature set to the shock temperature.  With the temperature and surface area, we can estimate the luminosity of SBO and with the size of the emitting region, we know our energy reservoir that, coupled with the luminosity, allows us to constrain the duration of the SBO burst.  These approximate solutions demonstrate that SBO is a direct signature of the shock velocity and stellar radius~\citep[for a review, see][]{2017hsn..book..967W}, providing a means of deeper understanding of supernova explosions and the stars which power them.  The advantage of this simple picture is that it allows a straightforward definition of SBO and shock cooling phases.

This SBO picture over-simplifies the early emission in a supernova explosion.  Studies over the past 15 years are just now starting to explore these complexities~\citep{2011AstL...37..194B,2013MNRAS.429.3181T,2017ApJ...850..133D,2020ApJ...898..123F}.  For example, we have assumed that the opacity is constant across the photon energy spectrum, but we know variations exist due to the density and temperature conditions~\citep[e.g.][]{2015ApJ...805...98B}.  Including the wavelength dependence and out of equilibrium effects will alter these results~\citep{1978ApJ...223L.109K,2007ApJ...664.1026W,2010ApJ...719..881S,2013ApJS..204...16F}.  In this case, focusing only on a narrow region at the front of the shock does not capture the observed SBO signal because this emission is produced across a much broader region.

Another complexity is that the velocity of SBO is not limited to the peak velocity of the forward blastwave within the star.  For example, as the shock front slows down, it creates a reverse shock that moves through the inner ejecta.  This means that the shock heated region can be much more extended than a simple forward-shock model assumes.  In addition, the stellar edge is poorly understood and its profile can dramatically affect the SBO signal~\citep{2015ApJ...805...98B,2017ApJ...845..103L}.  The stellar envelope and its edge can be inhomogeneous~\citep{2022ApJ...933..164G} causing the breakout to occur at different times at different parts of the star and drives a set of auxiliary shocks that further heat the shock.  Both of these effects extend the emission from SBO. Finally, explosion asymmetries from the engine also alter the SBO properties~\citep{2021MNRAS.508.5766I} and these asymmetries can also produce additional shocks.  All of these effects make it difficult to disentangle cooling flows from SBO effects.  As the extended cooling phase and true SBO are extremely difficult to disentangle in observations, we include them both in the model we describe in this paper. 

SBO effects can also be confused with other shock interactions that produce additional shock outbreak properties.  A complete study of SBO should include these broader set of emergent shock phenomena.  Examples of these studies include shock interactions in a clumpy wind medium~\citep[e.g.][]{2020ApJ...898..123F}, interactions with mass-loss just prior to collapse~\citep[e.g.][]{2017hsn..book..403S}, and shock interactions with companion stars~\citep[e.g.][]{2010ApJ...717..245K}.  These shock interactions are probes of the circumstellar medium and not of the stellar structure or explosion mechanism~\citep[for a review, see][]{2025arXiv250115702N}.  If these shocks occur at early times, they can be misinterpreted as SBO or contribute to the observed SBO signal. Table~\ref{tab:physicsmodels} lists the different SBO and broader shock interaction physics and their dominant effects on the SBO signal.

\begin{table*}
\centering
\begin{tabular}{|c|l|l|l|}
\hline\hline
 & \textbf{Process} & \textbf{Effects Beyond the Basic Model} &  \textbf{Model} \\
 &  &  &  \textbf{Implementation} \\
\hline
\multirow{6}{*}{\rotatebox[origin=c]{90}{\textbf{SBO Physics}}}
 & Forward Shock (basic model) & -- & \\
 & & & \\
 & Shock Acceleration & Higher velocities and emission temperatures &  $E(\Gamma \beta)$ \\
 & Stellar Asymmetries & Expands shock-heated region and broadens rise time & $E(\Gamma \beta)$ \\
 & SN Asymmetries & Expands shock-heated region and broadens rise time & $A_{\rm emit}(\Gamma \beta,t)$  \\
 & Reverse Shock & Expands shock-heated region & $E_0$ \\
 & Radiation-driven Shocks & Produces a low-mass shock leading the material shock & Not Included \\
\hline
\multirow{3}{*}{\rotatebox[origin=c]{90}{\textbf{Shock Interaction~}}}
 & & & \\
 & Clumpy Medium & Stochastic emission component & \\
 & & distributed heating broadening emission timescale & $\dot{E}_{\rm heat}$ \\
 & Shells & Prompt energy injection, sudden temperature rise & $\dot{E}_{\rm heat}$ \\
 & Companion Star & Prompt, localized energy injection & $\dot{E}_{\rm heat}$  \\
& & & \\
& & & \\
\hline\hline
\end{tabular}
\caption{Current analytical models only address the forward shock physics, ignoring many other processes that are known to occur.  Our implementation of the physics in our semi-analytical model includes a distribution of ejecta velocities, $E(\Gamma \beta)$, total energy reservoir, $E_0$, and emitting areas, $A_{\rm emit}(\Gamma \beta, t)$.  We also include a heating term, $\dot{E}_{\rm heat}$, to mimic late-time shock interactions. The details of our prescription are discussed in more detail in Sec.\ref{sec:SBOprop}).}
\label{tab:physicsmodels}
\end{table*}

With all of these effects included, SBO probes more than just the stellar radius and the shock velocity.  SBO observations probe the stellar profile at its edge (that can both help us better understand the mechanism behind stellar winds) and inhomogeneities in the stellar envelope.  SBO observations also probe asymmetries in the supernova explosion and the distribution of high velocity ejecta in that explosion, providing an ideal means to distinguish between jet-driven and neutrino-driven.  For the neutrino-driven model, we include both explosions revived quickly revived by neutrinos from the core and convection-enhanced models~\citep{1994ApJ...435..339H}.  In the simplest picture, massive stars go through a collapse/bounce phase where the bounce shock stalls through neutrino emission.  Neutrinos from low-mass stars (less than $\sim 12M_\odot$) can revive the shock.  At higher masses, convection above the proto-neutron star surface increases the efficiency of neutrino heating~\citep{1999ApJ...522..413F}.  For rapidly-rotating stars, primarily more massive stars likely collapsing to form black holes~\citep{2025ApJ...986..185F}, a jet-driven mechanism could drive explosions.  SBO measurements can probe the progenitors of these explosions to either confirm or complicate this basic picture of collapse-driven explosions.

SBO in gamma-ray bursts (GRBs) can provide complimentary information about the nature of supernovae produced by the jet-driven engine.  The first GRB model argued that SBO acceleration could produce ejecta with sufficiently high Lorentz factors that thermal 
emission from this shock, inverse-Compton up-scattered and Lorentz boosted into the observed gamma-rays~\citep{1968CaJPh..46..476C,1974ApJ...187..333C}.  But as it became more evident that GRBs were produced by relativistic jets, the SBO model lost its appeal.  However, even if SBO is not the dominant component to the GRB signal, it may still contribute to the GRB signal.  

Indeed, there are a growing number of papers arguing that a portion of the gamma-ray signal arises from thermal SBO emission~\citep{2011ApJ...727L..33G,2015ApJ...814...10G,2022ApJS..263...39M,2023ApJ...952L..42L}.  Comparing SBO models and their predicted spectra and spectral evolution to observations allow us to determine what fraction of the X-ray/gamma-ray signal can be explained by SBO.  We can then use the SBO portion of the signal to constrain properties of the ejecta.  Whether or not these signals are SBO, GRBs are ideal transients to study SBO events because wide-field gamma-ray instruments can be used to detect the initial burst of photons from the GRB jet and localized with X-rays.  With these jet observations providing a localization, the SBO of the stellar explosion associated with the GRB could be observed and could produce a GRB signal even without a highly-relativistic jet~\citep[e.g.][]{2006Natur.442.1008C}. 

But observations of these SBO events are difficult.  To date, the most convincing observation of SBO has been the serendipitous Swift observation of SN 2008D~\citep{2008Natur.453..469S}.  XMM may have detected a set of SN SBO in the X-rays~\citep{2020ApJ...896...39A}, but due to the high latency of identification follow-up observations were not performed and the classification of these signals is not known.  The transient community has developed a number of wide-field survey telescopes that are discovering supernovae at an unprecedented rate and these telescopes will allow both a higher rate of discovery and the ability to follow the cooling of the SBO~\citep[again, see the review by][]{2017hsn..book..967W}.  Wide-field X-ray satellites such as the Einstein Probe~\citep{2015arXiv150607735Y}, the recently launched BlackCAT mission~\citep{falcone2024blackcat}, and in-review AXIS Probe~\citep{2023SPIE12678E..1ER} along with wide-field ultraviolet detectors including forthcoming UltraSAT~\citep{2024ApJ...964...74S} and UVEX~\citep{2021arXiv211115608K} satellites are expected to dramatically increase the number of observed SBO events.

In this paper, we probe the broad set of SBO and shock emergence events by implementing a set of parameterized energy reservoirs, distributions of shock conditions, and emitting areas to better understand the range of shock emergence signatures in astrophysical transients. We do not explore additional emission mechanisms which act in some events (e.g. synchrotron radiation in GRBs).  These parameters are set by the nature of the explosion and we will differentiate between Type Ia SNe, Type II SNe, normal Type Ib/c SNe, broad-line Ic Supernovae (weak jet-driven explosions), as well as on and off-axis GRBs (strong jet-driven explosions).  Although the structure and properties of these outflows can vary dramatically, they can all be described by our SBO-model framework.  Section~\ref{sec:SBOprop} describes our framework for describing shock breakout and calculating its emission.  Using this framework, we discuss the properties and resultant emission for a wide range of models:  convection-driven core-collapse supernovae (Section~\ref{sec:exconvection}), jet-driven explosions including GRBs and broad line (Ic-BL) SNe (Section~\ref{sec:exjet}), type II SNe (Section~\ref{sec:exII}) and thermonuclear (type Ia) SNe (Section~\ref{sec:exIa}).  Section~\ref{sec:other} briefly discusses a few properties and observations to provide a flavor of their effects not fully studied in this paper.  Future work will systematically study these properties.  We compare our models to current observations in Section~\ref{sec:obscomp}.  Section~\ref{sec:projections} discusses the potential for current and future missions to observe shock breakout and we conclude with a summary of our key results.  


\section{SBO Properties}
\label{sec:SBOprop}

For the SBO emission model studied in this paper, we focus on the thermal Bremsstrahlung emission where the temperature is set by the shock heating.  The amount of shock heating is proportional to the kinetic energy in the shock, but as we shall see, the actual solution can be more complex than this.  For our simple parameterization, we will rely on the velocity distributions to provide us the initial temperature distributions in our shocks.   Coupled with the radius at SBO, we can use these velocity/temperature distributions to calculate the spectral emission from SBO.

Before we move to the model used in this paper, let's first understand SBO basics.  Simple SBO models assume that only the tip of the forward shock of thickness($\delta r$) is heated in the shock.  SBO occurs when the radiative velocity (relatively well approximated by the diffusion limit - $v_{\rm diff}$) exceeds the blastwave velocity ($v_{\rm bw}$).  The diffusive timescale is given by:
\begin{equation}
t_{\rm diff} = \lambda_{\rm mfp}/c (\delta r / \lambda_{\rm mfp})^2 = \delta r^2 \kappa \rho/c
\end{equation}
where $c$ is the speed of light and $\lambda_{\rm mfp} = 1/(\kappa \rho)$ is the mean free path where $\kappa$ is the opacity of the material (in ${\rm cm^2 \, g^{-1}}$) and $\rho$ is the density of the blastwave.  The diffusion velocity is then:
\begin{equation}
    v_{\rm diff} = \delta r/t_{\rm diff} = c/(\kappa \rho \delta r).
\end{equation}
By setting the radiation velocity to the blastwave velocity, we can determine the extent of $\delta r$:
\begin{equation}
    \delta r = (c/v_{\rm bw}) \kappa^{-1} \rho^{-1}
\end{equation}
and the mass enclosed ($M_{\rm bw}$) in this breakout region is roughly:
\begin{equation}
    M_{\rm bw} \sim \rho R_{\rm star}^2 \delta r = (c/v_{\rm bw}) \kappa^{-1} R_{\rm star}^2
\end{equation}
where $R_{\rm star}$ is the radius where SBO occurs, roughly the radius of the star.

The forward shock will accelerate as the blastwave propagates beyond the edge of the star.  The simplest picture for this shock acceleration is derived from the Taylor–von Neumann–Sedov similarity solution~\citep{1959sdmm.book.....S}:
\begin{equation}
    v_{\rm bw} \propto v_0 (t/t_0)^{(\alpha-3)/(5-\alpha)}
    \label{eq:sedov}
\end{equation}
where we assume the shock is moving through a density profile with a radial dependence of $r^{-\alpha}$.  If the radial profile of the stellar edge has a radial dependence dropping off quickly ($\alpha >3$), the shock will accelerate.  The front of the supernova blastwave can reach relativistic velocities (discussed in more detail below).  As this shock front decelerates after exiting the star and plowing through the stellar wind ($\alpha \approx 2$), the kinetic energy in the recently accelerated shock front is converted into thermal energy. 

The temperature ($T_{\rm SBO}$) of the shocked region can be estimated by assuming a fraction of the the kinetic energy is converted into thermal energy:
\begin{equation}
     a T_{\rm SBO}^4 =f_{\rm eff}/2 \rho v_{\rm bw}^2
    \label{eq:heatingsimple}
\end{equation}
where $a$ is the radiative constant and $f_{\rm eff}$ is the fraction of kinetic energy converted to thermal energy.   We use $f_{\rm eff}=0.1$ for most of our models which assumes that 10\% of the kinetic energy is converted into thermal energy which roughly matches radiation hydrodynamics calculations~\citep{2020ApJ...898..123F}.  With the temperature and surface area, we can estimate the luminosity of SBO and, with the size of the emitting region ($\delta r$), we know our energy reservoir that, coupled with the luminosity, allows us to calculate the duration of the SBO burst. 

To better capture the complexities of SBO, we broaden the emitting region to include a distribution of ejecta velocities.  The velocity distribution of the forward shock is not just a function of the energy of the explosion.  This distribution depends on a wide range of physical factors that must incorporate the shock acceleration and deceleration in the star as well as the immediate circumstellar medium.  One of the major effects on velocity distributions is the shock acceleration that occurs as the blastwave breaks out of its star (massive star or white dwarf).  This shock can accelerate dramatically as propagates down the steep density gradient at the edge of the stellar surface.  The simple Taylor–von Neumann–Sedov similarity solution (equation~\ref{eq:sedov}) indicates that the steeper the density gradient at the surface, the greater the acceleration.  \cite{2001ApJ...551..946T} included more detailed analyses of this shock acceleration determining the relativistic velocity distribution for a range of models \citep[eq. 37 of][]{2001ApJ...551..946T}):
\begin{equation}
        E (> \Gamma \beta) \propto (A E_{\rm exp}^{1/2})^{5.35 \gamma_p} F(\Gamma \beta)
        \label{eq:energydist}
\end{equation}
where $A$ is a coefficient based on the Taylor–von Neumann–Sedov similarity solution \citep[eq. 4 from][]{2001ApJ...551..946T}, $\gamma_p = 1 + 1/n$ where $n$ is the polytropic index for the star (which determines the density gradient), $E_{\rm exp}$ is the explosion energy, $\Gamma = \sqrt{1-\beta^2}$ is the Lorentz factor, $\beta = v/c$ (the ejecta velocity $v$ divided by the speed of light $c$, and $F(\Gamma \beta)$ is another function that depends on the density gradient \citep[eq. 38 from][]{2001ApJ...551..946T}).  \cite{2001ApJ...551..946T} studied a range of density profiles and explosion energies to determine energy distributions for different explosions including hydrogen giants (type II SNe), He/CO stars (Ib/c SNe) and white dwarfs (both accretion induce collapse and Ia explosions).  Their work found that, for specific conditions, the SBO of SN~1998bw could produce Lorentz factors that can, through synchrotron emission, explain the gamma-ray emission from this low luminosity GRB.  They argue that asymmetric explosions of some sort might be necessary for more energetic GRBs.

Calculating the ejecta velocity/Lorentz factor distributions has several uncertainties.  The analytic models do not include radiative processes and, in particular the radiation-material coupling.  Radiation can lead the material shock, reducing the energy in the shock and limiting the acceleration of the shock.  But it also also accelerates material ahead of the shock~\citep{2020ApJ...898..123F} that can alter the radiation flow entirely.  Unfortunately, high-fidelity radiation-hydrodynamical simulations are limited by numerical resolution.  These models can not reproduce the small amount of very-fast material seen in the analytic solutions, which is responsible for the high-energy emission.  Verification and validation studies of simulations (e.g. resolution studies, code comparisons to detailed analytic or semi-analytic models) are needed to understand this physics better.  

We can estimate the SBO light-curves and spectra using velocity profiles set by the conditions of different explosions.  As this high-velocity ejecta breaks out of the star, it decelerates in the wind ejecta, converting kinetic energy into thermal energy.  This thermal energy is then radiated, producing the observed SBO signal.  The standard SBO signal is calculated assuming all of the emission comes from the forward shock, neglecting the emission from the reverse shock and other secondary shocks due to the inhomogeneities in the shock.  Here we develop a more generic SBO emission model that better captures the physics observed in radiation hydrodynamics calculations.

For each transient, we expect a range of Lorentz factor distributions ($E(\Gamma \beta)$:  energy in the ejecta as a function of Lorentz factor) and emitting areas ($A(\Gamma \beta, t)$).  The distribution of energy as a function of Lorentz factor dictates both the peak photon energy and duration of the emission.  For our models, we assume a simple power-law distribution:
\begin{equation}
    E(\Gamma \beta) = E_0 (\Gamma)^{-p}
\end{equation}
where $p$ depends upon the progenitor and type of explosion (Table~\ref{tab:stellarmodels}) and $E_0$ is given by the energy in the explosion.

For the rest-frame emission, we assume a thermal Brehmstrahlung or blackbody:
\begin{equation}
L(\Gamma) = A(\Gamma \beta, t) \sigma T^4(\Gamma)
\end{equation}
where $\sigma = 5.67 \times 10^{-5} {\rm \, erg \, cm^{-2} \, K^{-4} \, s^{-1}}$ is the Stefan-Boltzmann constant and $T(\Gamma)$ is determined by assuming the kinetic energy is converted to thermal energy in the strong shock conditions.  In the Doppler-boosted frame, the initial temperature is given by:
\begin{equation}
T(\Gamma) = f_{\rm eff} 190 \Gamma^{0.5} (v_{\rm shock}/c)^{0.5} (\rho_{\Gamma}/10^{-10} {\rm g~ cm^{-3}})^{0.25} eV
\label{eq:temp}
\end{equation}
where $(v_{\Gamma}/c)$ is the ejecta velocity for the ejecta material with a Lorentz factor of $\Gamma$ in units of the speed of light and $\rho_{\Gamma}$ is the density of the material moving at Lorentz factor $\Gamma$.  As discussed in equation~\ref{eq:heatingsimple}, we set $f_{\rm eff} \approx 0.1$ for this paper.  Since we have a distribution of Lorentz factors, we have a distribution of initial temperatures for the different Lorentz factor regions.  Our calculation follows the evolution of each region of the ejecta as it cools through its radiative emission.  These models must also include Doppler-boosted (a.k.a. beaming) effects.

In our simplest model, we assume the emitting area ($A(\Gamma \beta, t)$) does not depend on $\Gamma$ and remains constant with time ($A(\Gamma \beta, t)= A $, a constant).  This will be the standard assumption in our study, unless noted otherwise.  In reality, we expect a distribution of emitting areas versus Lorentz factor and this distribution will alter the time-dependent spectra in our models.  In addition, especially for longer-duration emission events, the emitting area (i.e. position of the photosphere) will evolve with time.  We consider both of these effects in selected models of our study.  With this basic model, we just need to identify the distribution of velocities/Lorentz factors for different transients based on the explosion properties (e.g. velocity of the shock, jet and cocoon features) and the properties of the stellar edge/wind (to determine the amount of acceleration just before SBO).

Our approach in this paper is to use a parameterized set of models to mimic different stars (progenitors of Ia, Ib/c and II SNe) and explosive engines:  convective-driven paradigm~\citep{1994ApJ...435..339H} and jet-driven (a.k.a. collapsar) engines~\citep{1993ApJ...405..273W} for Ib/c and II SNe, and thermonuclear deflagration/detonations~\citep{1967ApJ...150..115F,2004PhRvL..92u1102G}.  Different explosions and different stellar density structures determine the amount of energy at different velocities and Lorentz factors.  With our shock heating model where we assume a fraction of the kinetic energy is converted to internal energy, this velocity distribution dictates a distribution of temperatures.  Coupled with the emitting surface area, a mass/volume of shock-heated material and an emergence timescale, we can calculate the Bremsstrahlung emission and cooling of this material to produce a SBO luminosity as a function of time.

For each transient, the distribution of velocities will differ either due to different properties of the engine or of the star and its circumstellar medium.  For supernovae driven by the convective engine, the parameterized solutions will be guided by the \cite{2001ApJ...551..946T} results but will include insight from the radiation hydrodynamics solutions and multi-dimensional calculations of supernova explosion.  Jet-driven explosions will include the properties of cocoon ejecta guided by collapsar models~\citep{2024arXiv240815973G}.  The difference between different supernovae correspond to the different properties of the star and its supernova explosion.  Table~\ref{tab:stellarmodels} shows both the properties of stellar progenitors of supernovae (C/O, He, H star radii and ejecta masses corresponding to Ic, Ib, and II supernovae respectively).  Unfortunately, the fidelity of current stellar models is incapable of truly modeling the edge of the star, although see recent results by~\citet{2022ApJ...929..156G}.  As such, we assume a broad range of properties and uncertainties remain large.  However, the basic trends are robust.  For example, the density profiles of C/O stars are expected to be steeper than He stars which, in turn, are steeper than H stars.  In addition, $E_{\rm in}$ is greatest for C/O stars and decreases with He and the H stars.  Similarly, the range of power-law coefficients is smallest for C/O stars, gradually increasing for He and then H stars.   

\begin{table*}
\begin{center}
\begin{tabular}{l|ccccc}
\hline\hline
Transient Type & Radius & M$_{\rm ejecta}$ & E$_{\rm in}$ & $p$ & $\beta \Gamma_{\rm max}$ \\
SN II & $2 - 20 \times 10^{12} {\rm \, cm}$ & $ 9 - 20 \, M_\odot$ & $0.001-0.1$ & $4-7$ & 0.1-1.5 \\
SN Ib & $2 - 5 \times 10^{10} {\rm \, cm}$ & $1.5 - 5.5 \, M_\odot$ & $0.001-0.1$ & $2-5$ & $0.5-10$ \\
SN Ic & $2-5 \times 10^9 {\rm \, cm}$ & $0.5-3.5\,M_\odot$ & $0.01-0.1$ & $2-4$ &  $0.5-10$ \\
SN Ic-BL jet & $2-5 \times 10^9 {\rm \, cm}$ & $10^{-5}-10^{-3}\,M_\odot$ & $0.1$ & $\sim 0-1$ &  $5-1000$ \\
SN Ic-jet off-set & $2-5 \times 10^9 {\rm \, cm}$ & $10^{-5}-10^{-3}\,M_\odot$ & $0.1$ & $\sim 0-1$ &  $5-1000^1$ \\
\hline
\end{tabular}
\caption{Properties of our different SBO models including the stellar progenitor (stellar properties based on pre-collapse models from \cite{2002RvMP...74.1015W}) and blastwave properties. The power law of the energy distribution ($p$) based on theory~\citep{2001ApJ...551..946T} listed here are in rough agreement with observed distributions~\citep{margutti2013signature}.  $^1$We study Ic-BL in two extremes, one where the jet is directed at the observer (with corresponding Lorentz factor boost) and one where the jet is perpendicular to the observer (no Doppler effects).  For Ia, a different formalism is used (see Section~\ref{sec:exIa}).}
\label{tab:stellarmodels}
\end{center}
\end{table*}

For themonuclear models (Type Ia SNe), we modify our prescription to include the expansion of the exploding whtie dwarf.  The next 4 sections cover Type Ib/c SNe, jet-driven Ib/c (both GRBs and Ic-BL SNe), type II SNe and thermonuclear (Ia) SNe.

\section{Ib/Ic Supernovae:  Convective Engine}
\label{sec:exconvection}

Type Ib/c supernovae occur in stars that have lost their hydrogen envelopes either through a common envelope event, extreme mass loss through winds or explosive mass ejections~\citep[for a review, see][]{2006ApJ...647.1269F}.  These stars tend to be more compact with radii less than $10^{11}$\,cm, or less than 1.5R$_\odot$ (compare this to $\sim 10^{12}-10^{13} {\rm \, cm}$ giant stars).  But the shock breakout radius can be beyond the stellar edge if the wind mass loss rate is so high that wind immediately above the star is optically thick.  Even in the strong wind case, the photosphere tends to be less than a few times $10^{11}$\,cm.  In this section, we study the expected shock interaction signals from these stars assuming a convective engine explosion.

Our simplest model assumes that the emitting area is constant both in time (photosphere stays near the stellar edge) and in Lorentz factor (that is, the emitting area for material at different Lorentz factors is the same).  This simple model provides a first approximation of supernova explosions where we can study the emission for a range of emitting areas, energies and distribution of energies as a function of Lorentz factor/temperature.  For this emission, we only study the material with velocities above $\sim 3 \times 10^9 {\rm \, cm \, s^{-1}}$ ($\Gamma \beta > 0.1$).  These different parameters correspond to different supernova types (see Table~\ref{tab:stellarmodels}).  

\begin{figure}
    \includegraphics[width=0.5\textwidth]{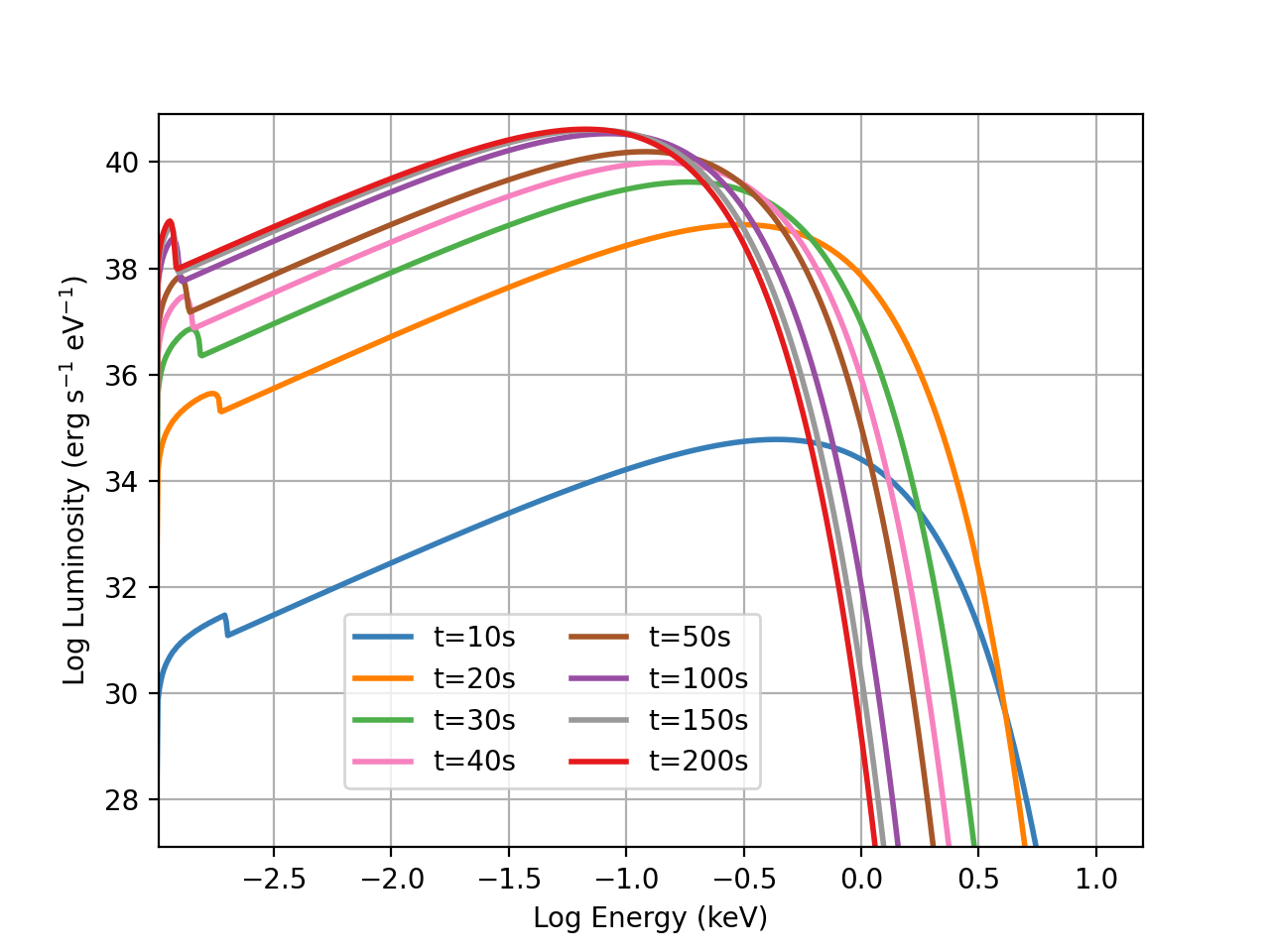}
    \includegraphics[width=0.5\textwidth]{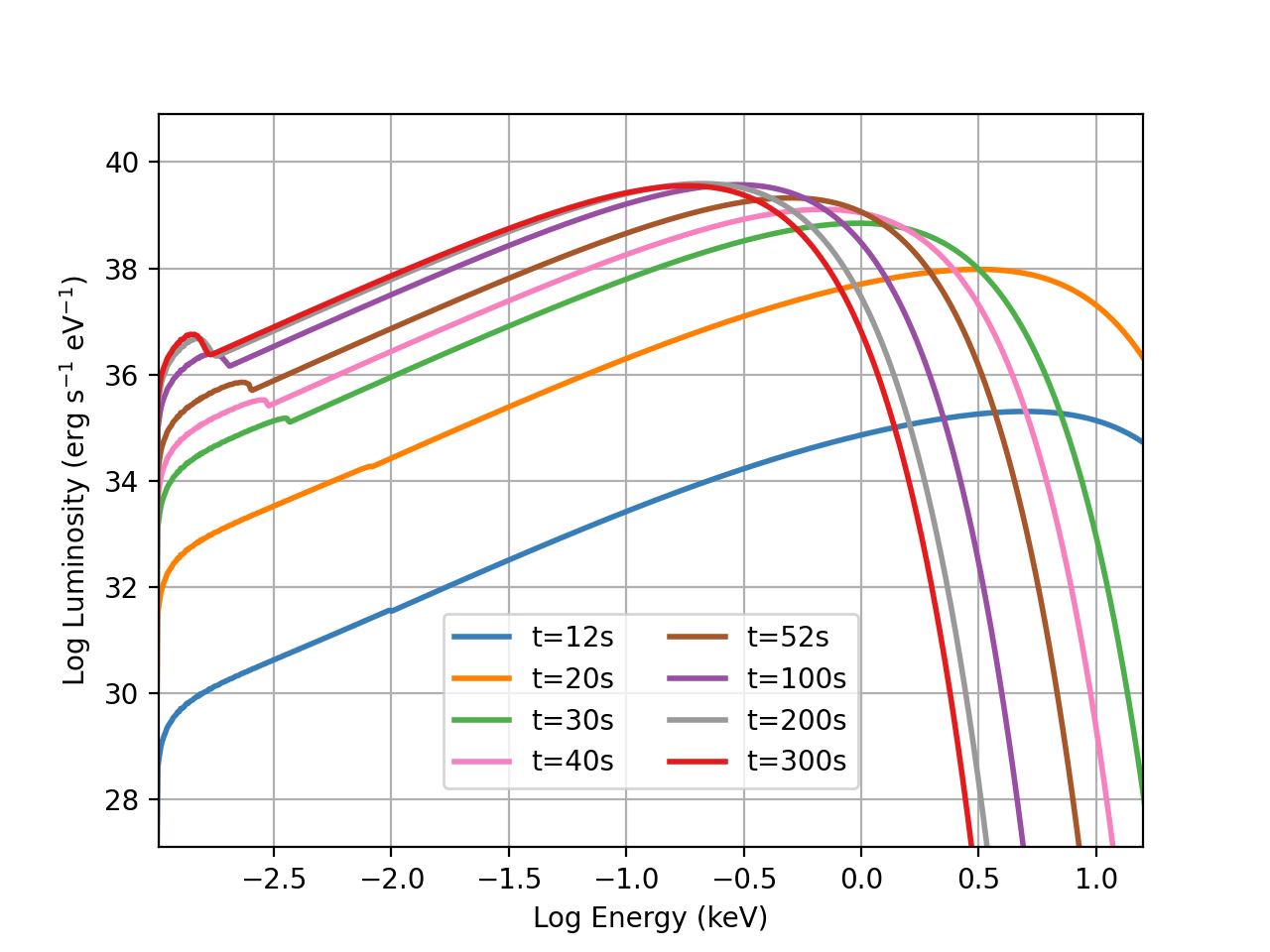}
    \caption{Spectra as a function of time from the initial shock breakout from two of our SBO models for Ibc supernovae driven by convective engines: (top) $E_{\rm in} = 10^{48} {\rm erg}$, Area $=10^{24} {\rm \, cm^2}$, $p=6$, $\Gamma_{\rm max}=2$; (bottom) $E_{\rm in} = 10^{48} {\rm erg}$, Area $=10^{24} {\rm \, cm^2}$, $p=4$, $\Gamma_{\rm max}=10$.  These models initially peak in the X-rays (a few keV) but the peak photon energy quickly decreases to a few hundred eV or less.  Covering the emission from 1 eV up to 10\,keV is required to study all aspects of this emission.} 
    \label{fig:stripped_spec}
\end{figure}

For our stripped-star supernova studies with constant radius/surface area, we consider 3 energies in high-velocity ejecta ($\Gamma v/c > 0.1$):  $10^{48}, 10^{49}, 10^{50} \, {\rm erg}$, 3 power laws for the energy distribution (equation ~\ref{eq:energydist}, $p=2,4,6$), and 5 emitting areas ($10^{20}-10^{24} \, {\rm cm^2}$).  For most of these models, we set the peak Lorentz factor ($\Gamma_{\rm max}$) to 10. But we include a few models with a cut-off in the peak Lorentz Factor of the ejecta ($\Gamma_{\rm max} = 1.1-50$).  Figure~\ref{fig:stripped_spec} shows the spectra from 2 of these models:  one showing a higher power law value ($p=6$) and a relatively low peak Lorentz factor ($\Gamma_{\rm max} = 2$), another with $p=4$ and our standard $\Gamma_{\rm max} = 10$ Lorentz factor.  In both models, the photon energy of the emission initially peaks above a few keV but quickly drops below a few hundred eV after 300s.

\begin{figure}
    \includegraphics[width=0.5\textwidth]{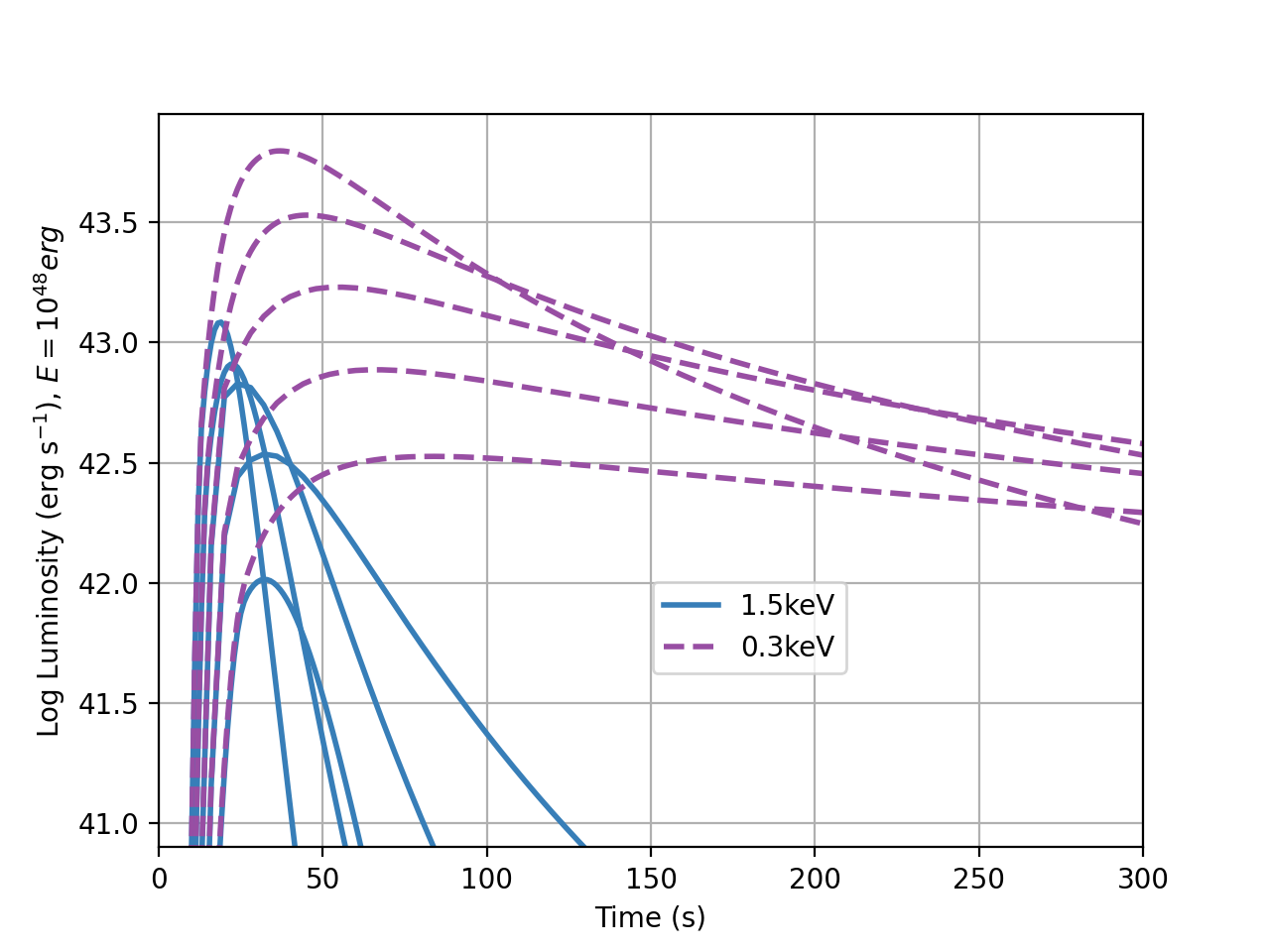}
    \includegraphics[width=0.5\textwidth]{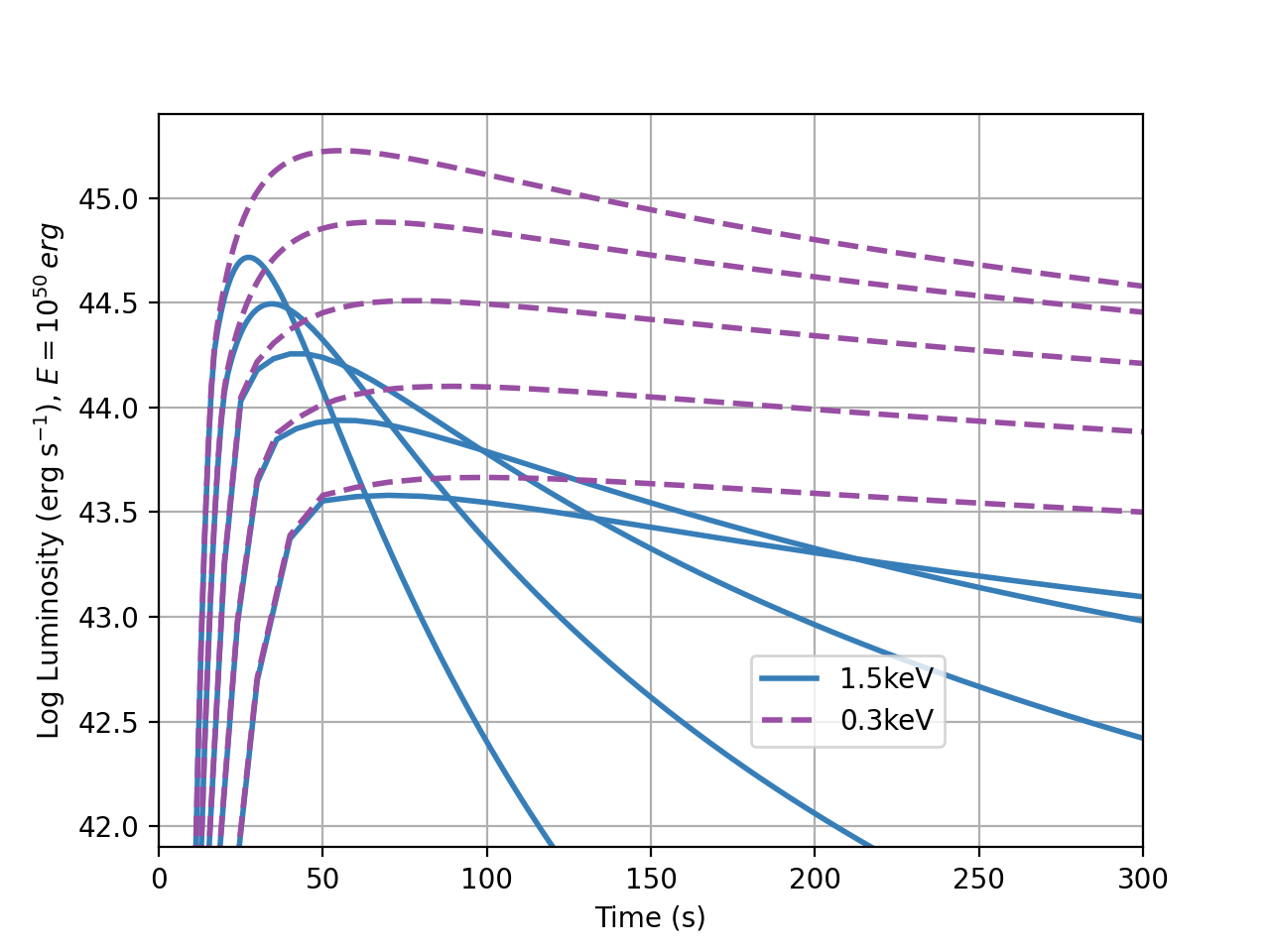}
    \caption{X-ray Luminosity above 0.3\,keV (dashed) and above 1.5\,kev (solid) for 2 model suites with total fast-ejecta energies ($\beta \Gamma >0.1$) of $10^{48}$ (top) and $10^{50} \, {\rm erg \, s^{-1}}$ (bottom) and $p=4$.  For each of these energies, we study the light curves for 5 emitting areas:  $10^{20},10^{21},10^{22},10^{23},10^{24} {\rm \, cm^2}$.  The peak luminosity scales with total energy (although not exactly linearly).  Higher emitting areas lead to brighter peaks of shorter duration.  A diverse set of light-curves can be produced varying these 2 parameters.} 
    \label{fig:flxray_area}
\end{figure}

Figure~\ref{fig:flxray_area} shows the dependence of the X-ray luminosity of photons above 0.3\,keV for models with different energies and emitting areas.  In comparison, we show the same light-curves if the limiting photon energy is 1.5\,keV.  At 0.3\,keV, we expect many systems with luminosities exceeding $10^{43} {\rm \, erg \, s^{-1}}$ at peak with long-lived X-ray emission that remains within 1/10th of peak luminosity for over 1000~s.  The luminosity at higher energies peaks roughly a factor of 3 lower and the duration is often much shorter falling below 1/10th of peak in 10-100s for many models.

\begin{figure}
    \includegraphics[width=0.5\textwidth]{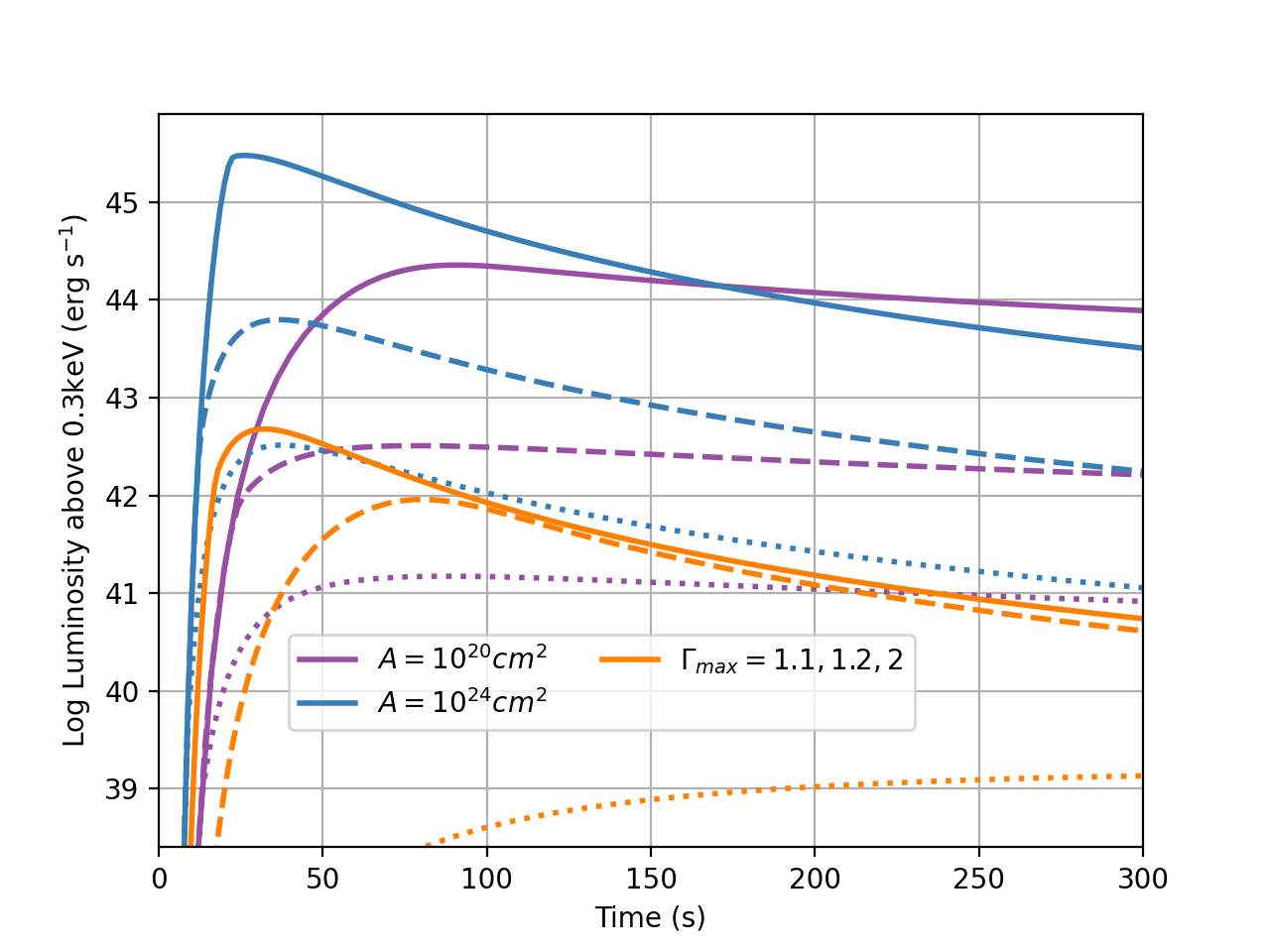}
    \caption{X-ray luminosity for 2 convective-engine driven Ibc model suites with emitting areas of $10^{20}{\rm \, cm^2}$ and $10^{24} {\rm \, cm^2}$ with a total energy of $10^{48}$~erg.  The larger emitting area leads emission that peaks at a higher luminosity but evolves more quickly than the lower emitting area. The peak Lorentz factor is 10, the total energy with velocities $\beta \Gamma > 0.1$ and $\Gamma_{\rm max}$ are kept the same while varying the power-law index of the distribution of Lorentz factor:  $p$ = 2 (solid), 4 (dashed), 6 (dotted).  To study the role of this peak Lorentz factor, we include 3 additional $p$=6 models varying this peak value as shown in orange:  $\Gamma_{\rm max}$ = 2 (solid), 1.2 (dashed), 1.1 (dotted).} 
    \label{fig:flxray_slope}
\end{figure}

For our base convective-driven Ib/c SN models, we assumed a peak Lorentz factor of 10 and a fairly steep power law ($p=6$) for the distribution of Lorentz factors.  But uncertainties in the shock acceleration as it breaks out of the star make it difficult to pin down the Lorentz factor distribution.  To understand the importance of this acceleration, we compare models with different power-law slopes of the Lorentz-factor distribution (Figure~\ref{fig:flxray_slope}).  We also study a range of models with caps on the maximum Lorentz factor from 1.1 to our standard value of 10.  Varying the slope from 2 to 6 alters the peak luminosity by more than 3 orders of magnitude.  Capping the maximum Lorentz factor can have a similarly dramatic effect.  Observations of the X-ray flux will allow us to place constraints on the distribution of ejecta Lorentz factors, providing direct insight into the outcome of these competing physics in SBO.

Although the characteristics of shock breakout from convective-engine core-collapse supernovae vary considerably, there are a few trends that persist throughout the models.  For all of these models, the emission lies in photon energies between 0.1-10\,keV.  Photons with energies above 10\,keV also have a much shorter duration than X-ray photons below 10\,keV.  All else being equal, the models with high Lorentz factors will produce more energetic photons.  The total energy in the fast ejecta and the area of the photosphere both dictate the peak emission, but peak emission and duration can differentiate the two, allowing astronomers to determine both with full signals.  Combined with a measurement of the energy distribution, the characteristics of the emerging shock can be probed.

\section{Jet-Driven Explosions:  Ic-BL SNe and GRBs}
\label{sec:exjet}

Another class of explosions can be produced by rotating massive stars forming disks that drive jets (e.g. long-duration GRBs and Ic-BL SNe).  A broad range of models have been performed studying the propagation of collapsar jets through their host stars beginning with early models by \cite{1999ApJ...524..262M} and extending to current research with increasingly sophisticated models~\citep[for a review, see][]{2024arXiv240815973G}.  Although some disk models can produce strong jets, resolution constraints in the magneto-hydrodynamic calculations limit the accuracy of these models.  As the jet propagates through the star, it imparts energy onto the star, both through formation of an accelerated shear region (a.k.a. ``cocoon'') and, even without shock acceleration as the blastwave breaks out of the star, we expect a broad range of Lorentz factors in the ejecta material both from theoretical models~\citep{1999ApJ...524..262M,2021MNRAS.500.3511G,2022A&A...668A..66J,2023MNRAS.518.5145U,2025MNRAS.538.1247U} and observations~\cite{2018A&A...609A.112G,2022Galax..10...93S}. Note that in this section we do not discriminate between jet and supernova emitting regions, reasonably treating them as connected regions arising from the same underlying ejection process.

The peak Lorentz factor depends both on the power in the jet and the amount of baryonic material swept up in the jet.  The range of Lorentz factors will depend on the amount of energy injected into the star either from the jet (including the cocoon) or disk wind.  In the turbulent shear layer, calculations of the cocoon show that Lorentz factors up to 10 can be achieved~\citep{2024arXiv240815973G}.  If the shear drives extensive mixing, it is possible to mix so much mass into the jet that this baryon contamination will choke the jet.  With distinct ejecta regions (jet, cocoon, broader ejecta), we can understand the variety of jet-driven supernovae:  e.g. low-luminosity GRBs may arise from an explosion dominated by the cocoon region~\citep{2012ApJ...747...88N} and Ic-BL supernovae with no GRB (low-luminosity or otherwise) may require choked jets~\citep{2012ApJ...749..110B,2019ApJ...871L..25P} and even weak cocoon ejecta. It is possible that all Ic-BL supernova seen, overwhelmingly of type Ic, arise from jetted supernova engines \citep{2025ApJ...986..185F}.

\begin{figure}
    \includegraphics[width=0.5\textwidth]{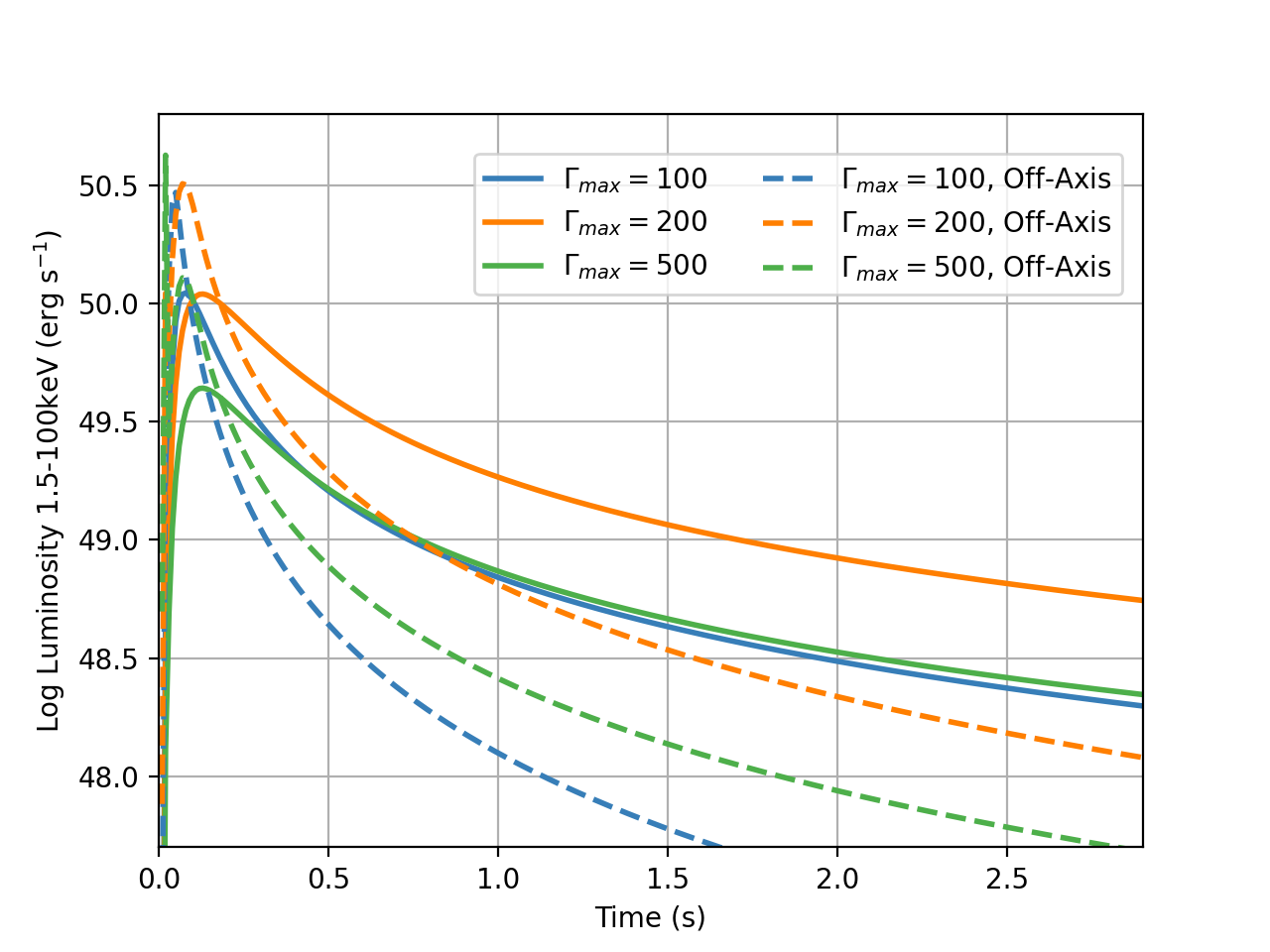}
    \caption{Luminosity (erg \, s$^{-1}$) between 1.5-100\,keV versus time for strong jet ($\Gamma = 100, 200, 500$) explosions both on-axis (solid lines) and (maximally) off-axis (dashed lines).  For the on-axis models, the high Lorentz factor Doppler shifts much of the emission to energies above 100\,keV at early times.  As such, the off-axis models are brighter in the X-ray at early times.  The integrated luminosities will also be higher.}
    \label{fig:GRBjetlum}
\end{figure}

\begin{figure}
    \includegraphics[width=0.5\textwidth]{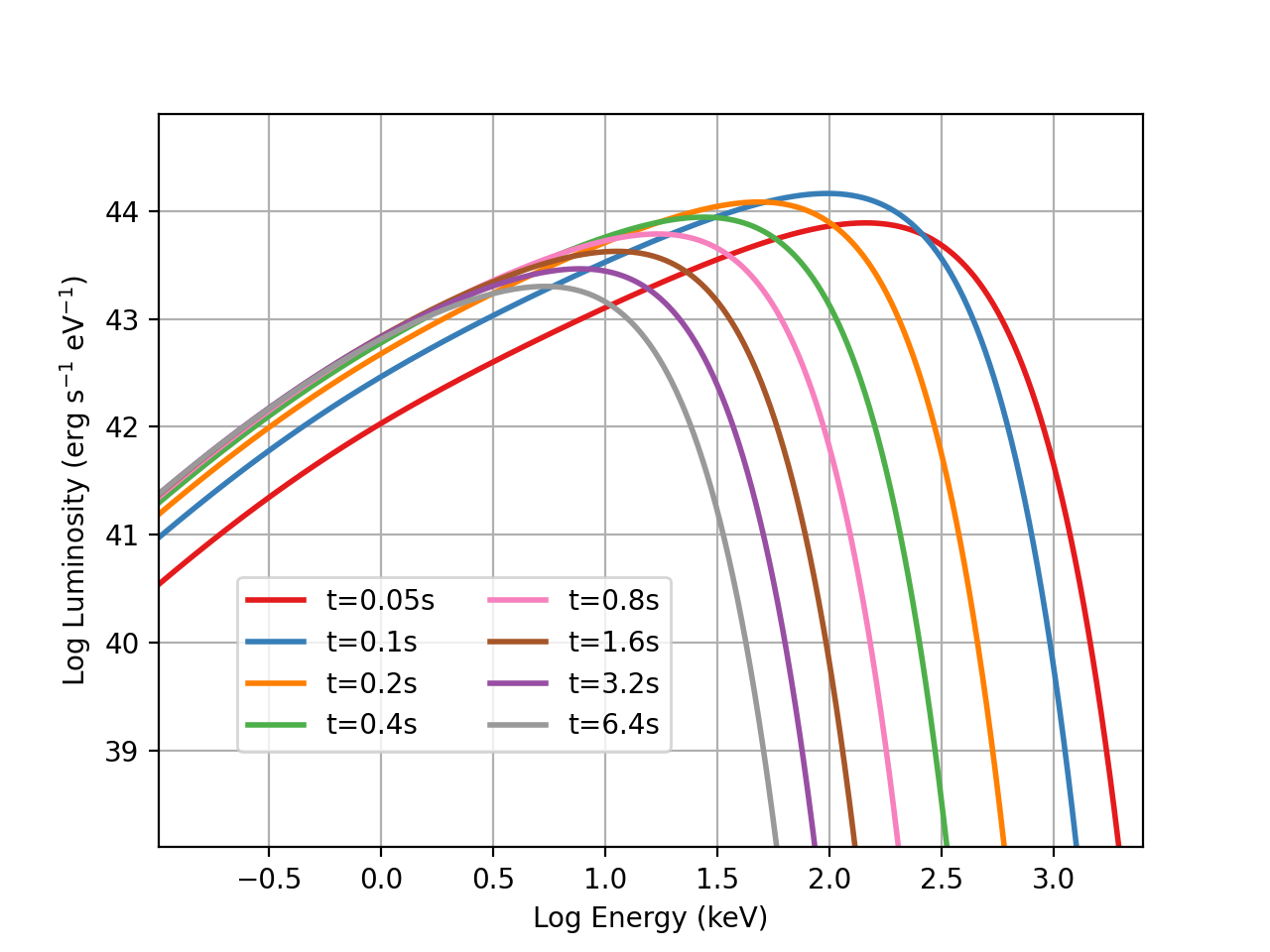}
    \includegraphics[width=0.5\textwidth]{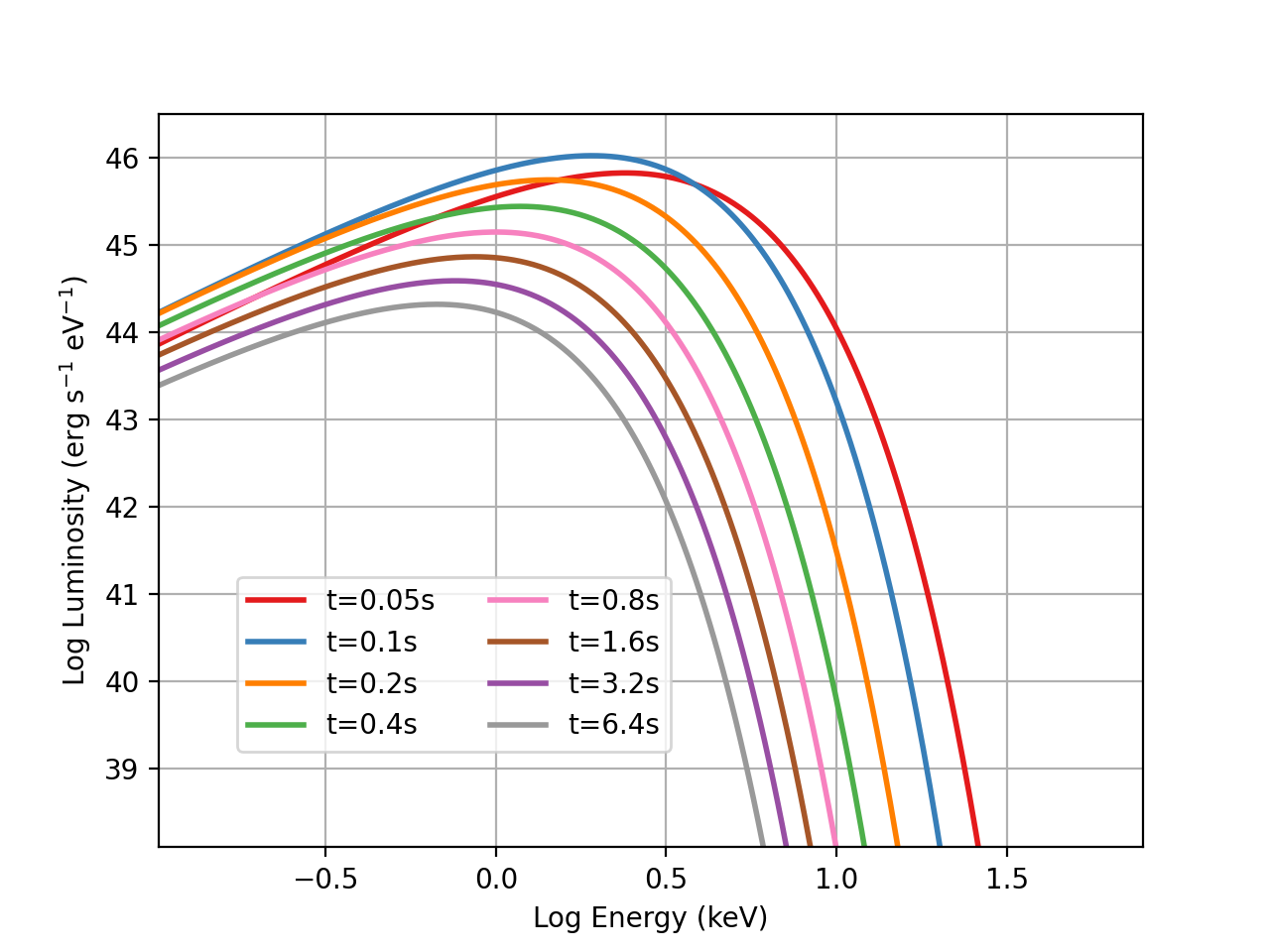}
    \caption{Spectra for a series of snapshots in time from our $\Gamma=200$ model both on-axis (top) and off-axis (bottom).  For the on-axis model, the emitted photons are initially above a few hundred keV, including some emission above 1\,MeV.  But the ejecta quickly cools and, within 0.5\,s, the photon energy has already dropped below a few hundred keV.  Without the Lorentz boost, the emission from the  off-axis model is below 30\,keV.}
    \label{fig:GRBspecevol}
\end{figure}

It is likely that these jet driven explosions have Lorentz factors that have a peak value that ranges from 10 to a few hundred (if not greater).  In addition, the diversity argues that the distribution of Lorentz factors in the ejecta is not described by a simple power law.  Nonetheless, we assume that the distribution of Lorentz factors will be much flatter than those expected from convective-engine driven explosion.  For our models we will assume a value $p=0-2$, in rough agreement with observations~\citep{margutti2013signature}.  For luminous GRBs, we expect high Lorentz factors ranging from 100-500.  Figure~\ref{fig:GRBjetlum} shows the X-ray luminosity from these high Lorentz-factor jets.  For our on-axis jets, the $\gamma$-ray SBO signal can exceed $10^{50} {\rm \, erg \, s^{-1}}$ but this luminosity drops off within the first few tenths of a second as the shock cools (Figure~\ref{fig:GRBspecevol}).

Our off-axis models assume the observer is $90^{\circ}$ off of the jet axis.  Although the Lorentz factor distributions are the same as our on-axis models (and hence the temperature distribution is the same), without the maximal Doppler boost, the emission from the jets is in the X-ray band, not $\gamma$-ray band.  Because of this, these off-axis simulations are actually brighter in X-rays than the on-axis models (Figure~\ref{fig:GRBjetlum}).  These models cool quickly.  The emission quickly drops below most X-ray detector sensitivities and the observed the observed X-ray emission is short-lived.  

\begin{figure}
    \includegraphics[width=0.5\textwidth]{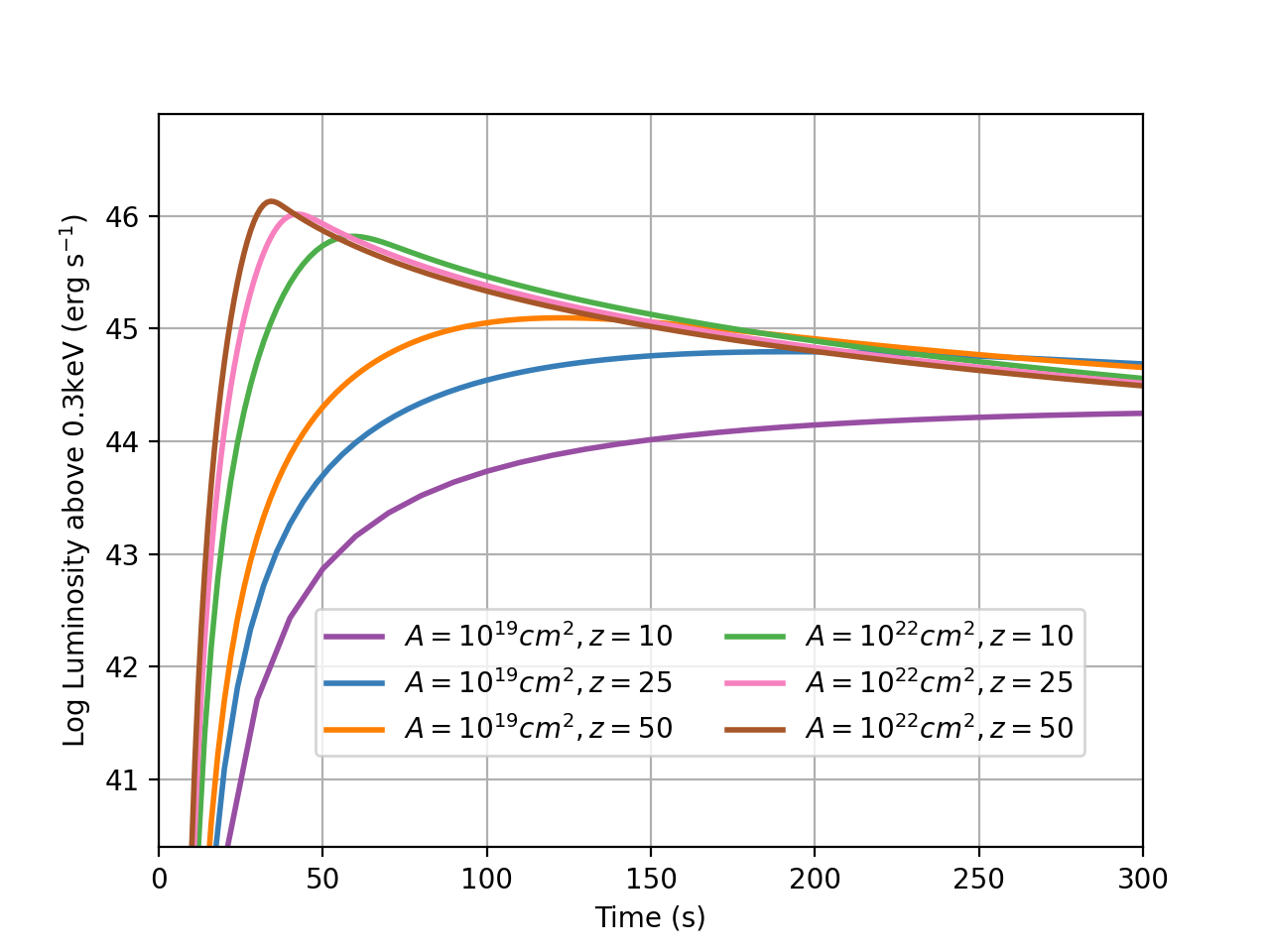}
    \caption{High energy fluxes for our weak jet-driven supernovae using a relatively flat ($p=1$) Lorentz factor distribution with $10^{48}$\,erg of energy with velocities $\beta \Gamma > 0.1$.  These models are brighter than our convective-driven engine models.  As we see in Figure~\ref{fig:fspec_gam}, the peak photon energies are also much higher.} 
    \label{fig:flxray_gam}
\end{figure}

We also study a set of partially failed jet models with maximal Lorentz factors lying from 10-50.  These Lorentz factors are not high enough to produce bright GRBs, but they could produce low-luminosity GRBs and failed-jet Ic-BL SNe.  
The high-energy flux for on-axis models with different emitting areas and modest peak Lorentz factors is shown in Figure~\ref{fig:flxray_gam}. 

\begin{figure}
    \includegraphics[width=0.45\textwidth]{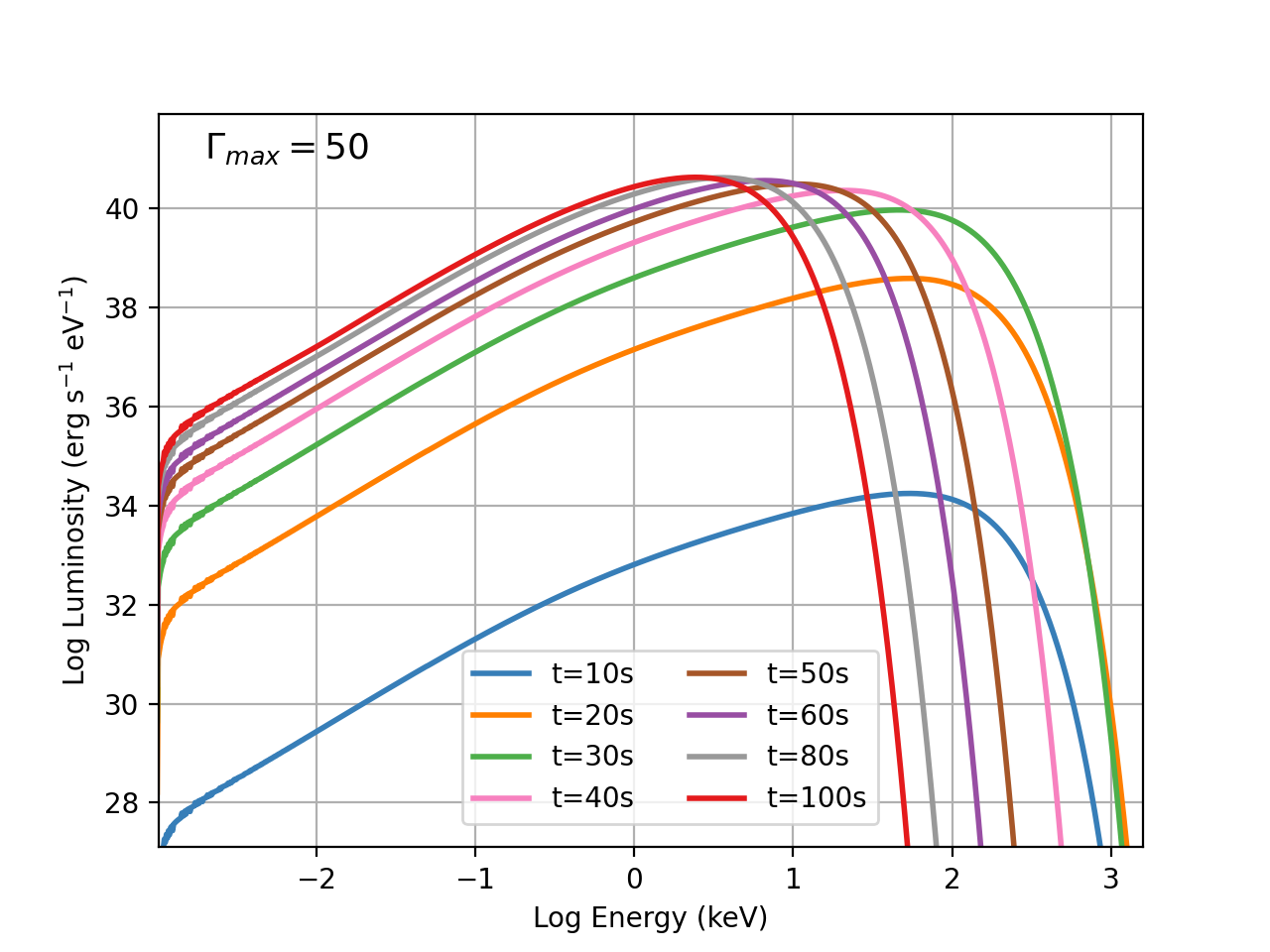}
    \includegraphics[width=0.45\textwidth]{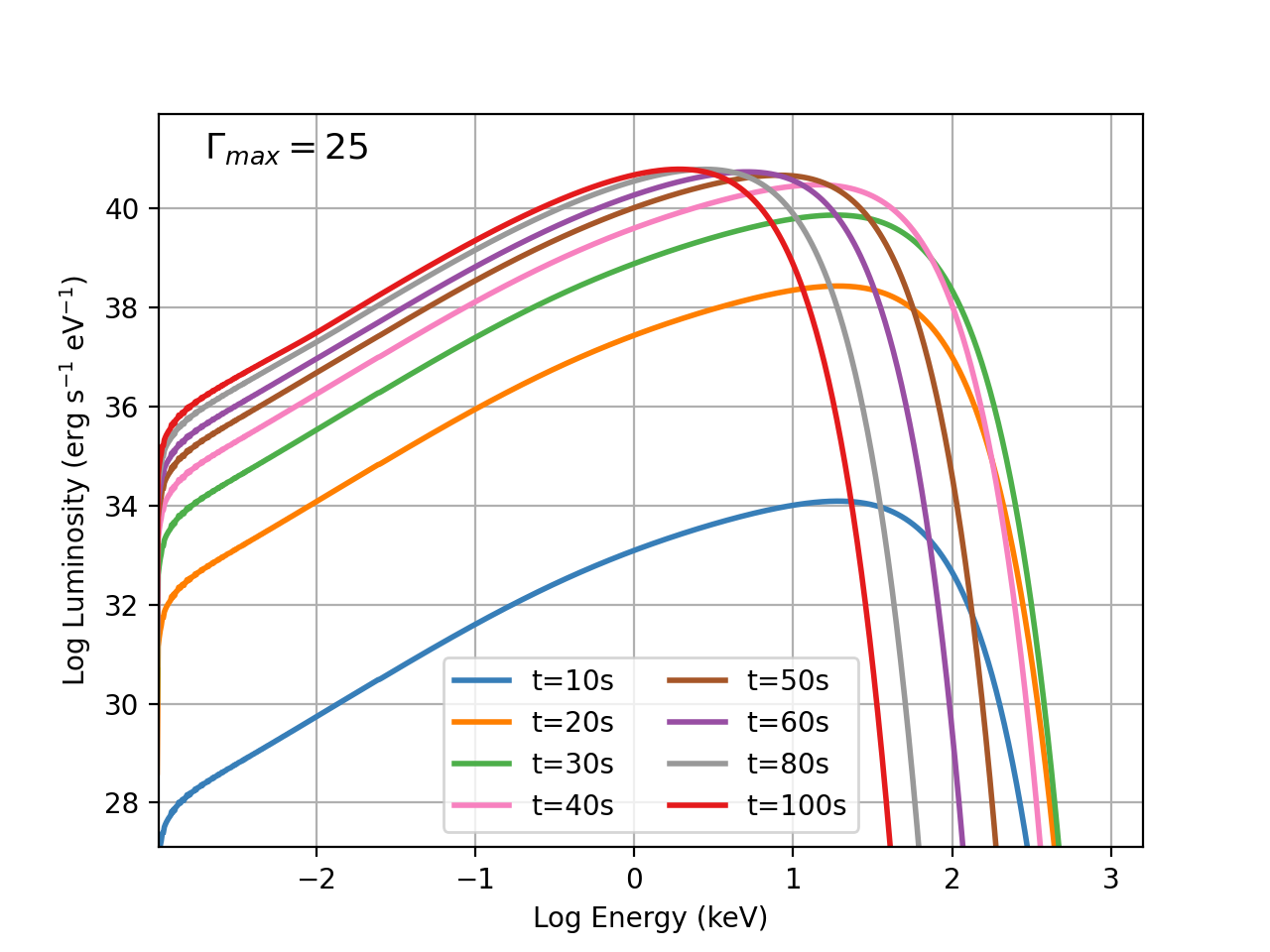}
    \includegraphics[width=0.45\textwidth]{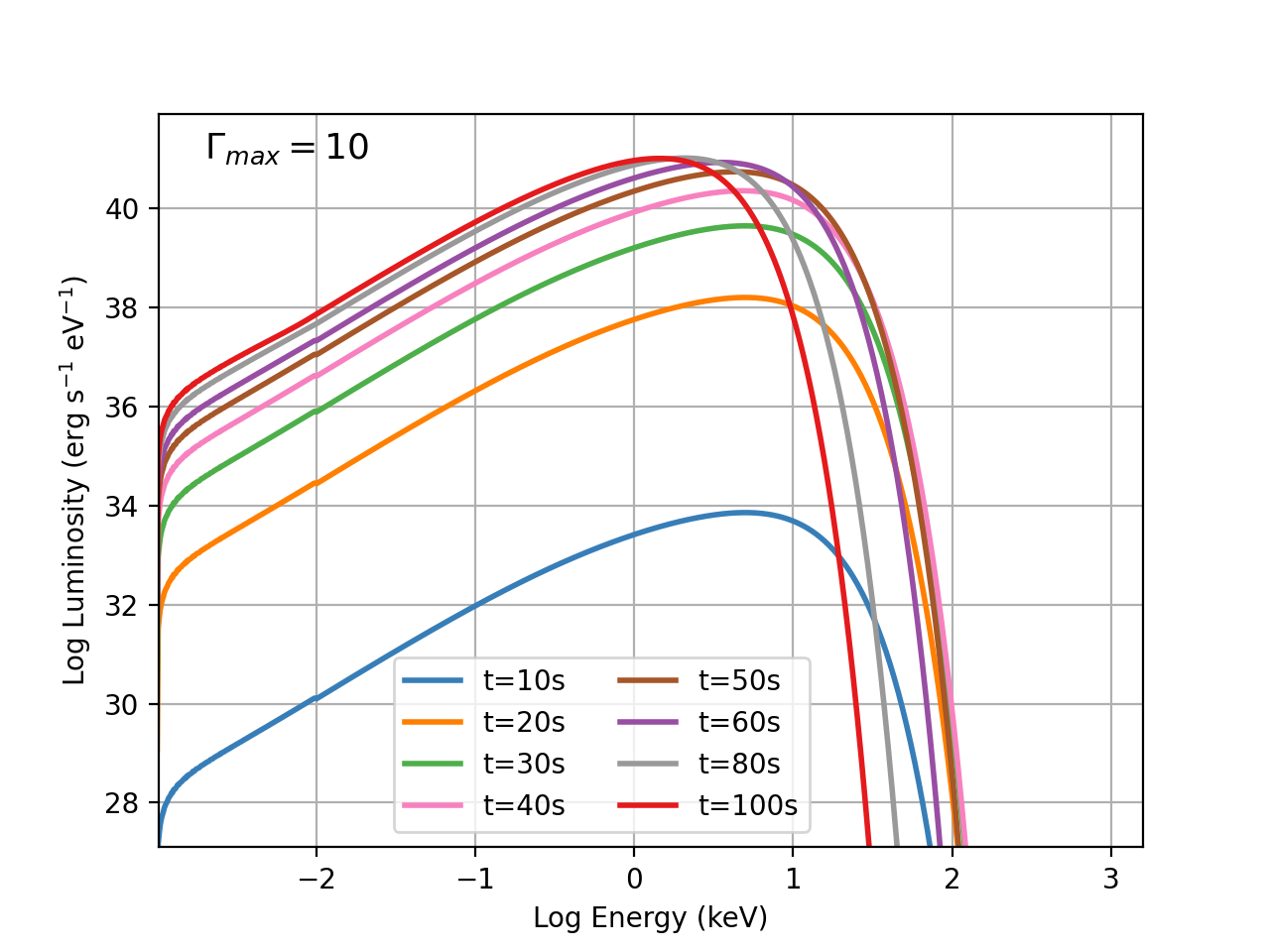}
    \caption{Time dependent spectra from the SBO of a failed jet or cocoon ejecta assuming a shallow slope in the Lorentz factor distribution ($p=1$) and a maximum Lorentz factor of 50 (top), 25 (middle) or 10 (bottom).  For these models, we assume an emitting area of $10^{22}$\, cm and a total energy with $\beta \Gamma>0.1$ set to $10^{48}$\,erg.  For such models, the thermal emission for a $\Gamma_{\rm max}=50$ jet initially produces emission with photon energies peaking around 50\,keV.  In 100\,s, it cools and the peak photon energy lies below 10\,keV.  The peak photon energy for the $\Gamma_{\rm max}=25$ model is closer to 20\,keV.} 
    \label{fig:fspec_gam}
\end{figure}

The spectra from these models is highly sensitive to the maximum Lorentz factor of the ejecta.   Figure~\ref{fig:fspec_gam} shows the spectra as a function of time for a shallow slope $p=1$ calculation with maximum Lorentz factors of 25 and 50.  In these cases, the emission peaks initially in the gamma-rays.  Coarse spectral resolution in the X-rays would allow us to place constraints on the Lorentz factor, allowing us to prove the existence and strength of the jet.  Such observations would allow us to differentiate jet-driven versus convective-engine driven models for Ib/c and Ic-broad-line supernovae.

\begin{figure}
    \includegraphics[width=0.45\textwidth]{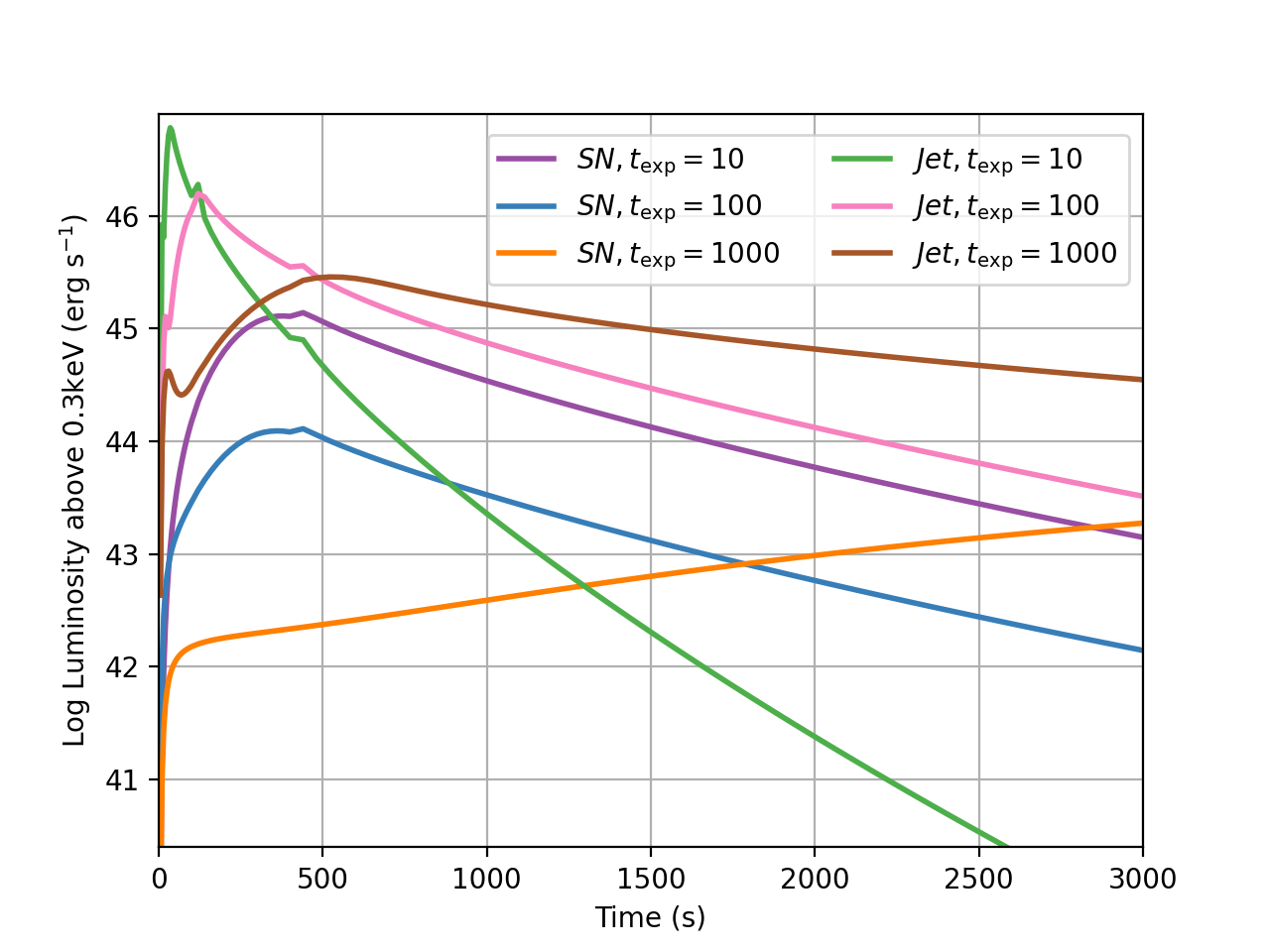}
    \caption{X-ray emission versus time for expanding photospheres of two different models characteristic of convective (``SN'': $p=2, \, \Gamma_{\rm max}=2$) and jet-driven ($p=1, \, \Gamma_{\rm max}=20$) engines for type Ic supernovae.  For both models, we assume the initial emitting area is small ($10^{19} \, {\rm cm^2}$), comparable to the surface of a compact C/O core with low wind mass-loss.  We then expand this area using a range of expansion timescales ($t_{\rm exp}$) for equation~\ref{eq:expansion}.} 
    \label{fig:expansion}
\end{figure}

In these models, we have assumed that the emitting area has remained constant throughout the explosion.  As with any ejecta-driven transient, the photosphere will initially expand and then, as the ejecta becomes diffuse, move inward in mass, and ultimately, radius~\citep{doi:10.1142/S0218271825400024}.  For the bulk of the supernova ejecta, the stagnation or inward motion in radius of the photosphere occurs after 20\,d and is ultimately the cause of the drop in the supernovae light-curve.  For the low-masses in the relativistic ejecta producing SBO X-rays, the photosphere often remains very close to the stellar edge, justifying our choice of a constant area for the photosphere. However, this photosphere can initially move outward.  To understand this effect, we also have a set of runs where the photosphere moves outward with time and the emitting area ($A_{\rm emit})$ increases with this outward motion:
\begin{equation}
    A_{\rm emit} = A_{\rm emit(0)} (1+(t/t_{\rm exp})^2)
    \label{eq:expansion}
\end{equation}
where $t$ is the time and $t_{\rm exp}$ is the expansion time that we range from 10 to 1000\,s.  Figure~\ref{fig:expansion} compares two different explosion models, a convective engine ($p=2, \, \Gamma_{\rm max}=2$) and a jet-driven engine ($p=1, \, \Gamma_{\rm max}=20$).  For both models, we assume an initial emitting area set to a fraction of the surface area of the compact C/O core ($\sim 10^{19} \, {\rm cm}$ which corresponds to a hot spot in the ejecta near the jet axis) with sufficiently low wind that the initial breakout occurs in the star, not the wind.  We then expand the emitting area within our expansion time range of 10-1000\,s.  For these compact stars, the expansion can be sufficiently fast to dramatically alter the light curves.  Detailed explosion models can be used to estimate the true nature of the photospheric expansion.

Even for our high Lorentz-factor jet models, the peak shock temperatures are typically at a few tens of keV.  But the Lorentz-boosted photons can peak in the MeV range.    As this radiation interacts with the circumstellar medium just above the shock, pair-production is expected and may ultimately affect the emission~\citep{1976ApJS...32..233W}.  Off-axis observations will peak in the X-rays as will many failed jet scenarios.  The peak emission energy depends both on the peak Lorentz factor and the amount of energy in the material at this peak Lorentz factor.  In many cases, the highest Lorentz factor material cools before the peak in the light-curve.  Photospheric expansion further complicates this picture.  The energy of the photons and photon index for the high-energy emission at early energies can provide direct insight into the peak Lorentz factor and the Lorentz factor distribution.

\section{Type II Supernovae}
\label{sec:exII}

\begin{figure}
    \includegraphics[width=0.5\textwidth]{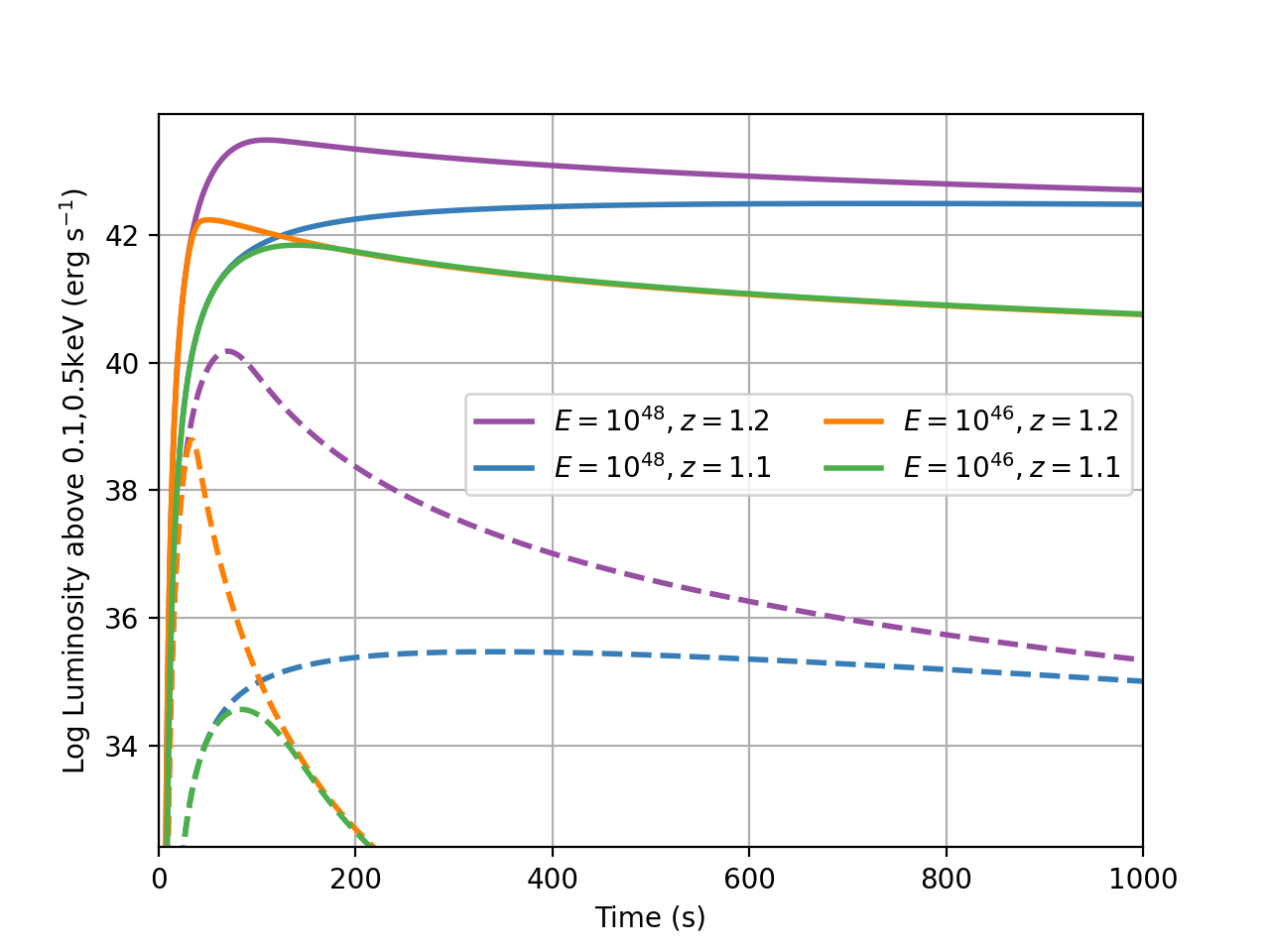}
    \caption{X-ray luminosities for 4 type II SBO models with 2 different peak Lorentz factors:  $\Gamma_{\rm max} = 1.1, 1.2$ with a total energy of $10^{46}, \, 10^{48} {\rm \, erg}$ with velocities $\beta \Gamma > 0.1$.  The SBO of type II SNe can be bright if we can detect the emission down to 0.1\,keV (solid lines).  But the peak luminosity above 0.5\,keV (dashed lines) is 3--7 orders of magnitude dimmer.} 
    \label{fig:xrayII}
\end{figure}

For our type II supernovae, we focus only on the fast ejecta produced as the (convective-engine driven) supernova blastwave propagates through the stellar boundary.  The SBO signal depends sensitively on the amount of material that accelerates as the supernova blastwave breaks out of the star.  This, in turn, depends on the sharpness of the boundary between star and stellar wind.  In red supergiants, the star/wind boundary could well be more gradual than is predicted by the outer density of stellar models and the expected wind densities~\citep{2022ApJ...929..156G}.  As such, it is difficult to predict the SBO emission.  For our models, we assume $10^{-7}-10^{-5}\,M_\odot$ of ejecta is accelerated to velocities above 10\% the speed of light.  These masses correspond to an energy at these high velocities of roughly $10^{46}-10^{48}\,{\rm erg}$.  Figure~\ref{fig:xrayII} shows the X-ray luminosity as a function of time for these two energies assuming a peak velocity of $0.4-0.55$ times the speed of light.  The peak luminosity in X-rays of type II SBO events are much lower than those of Ib/c supernovae.  Even more critical, these events will not be detectable by instruments with limiting low-energy thresholds above 0.5-1\,keV. Detections then require lower low-energy thresholds; however, here extinction begins to become a major factor.

\begin{figure}
    \includegraphics[width=0.5\textwidth]{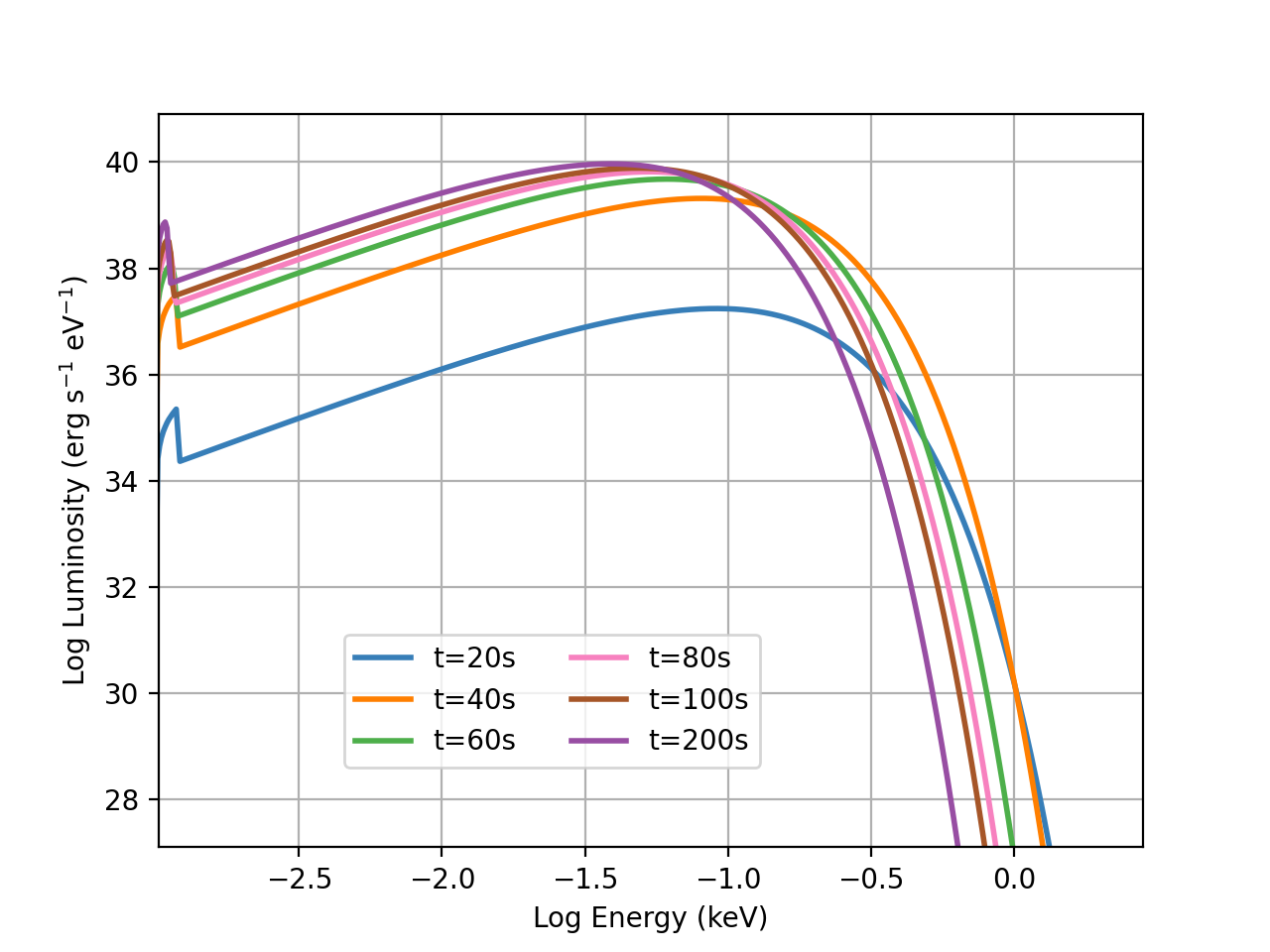}
    \caption{Spectra as a function of time of our Type II SN model with a total energy of $10^{46} \, {\rm erg}$ with velocities $\beta \Gamma>0.1$ and a maximum Lorentz Factor $\Gamma_{\rm max}=1.2$.  Most of the emission from these supernovae will have photon energies below 0.1\,keV, peaking between 30--100 eV.} 
    \label{fig:specII}
\end{figure}

Figure~\ref{fig:specII} shows the spectra for our $10^{46} \, {\rm erg}$, $ \Gamma_{\rm max}=1.2$ model as a function of time.  The peak energy in the emission quickly falls below 0.1\,keV.  But, at the same time, the peak emission lies above the far ultraviolet (UV observations will be discussed in more detail in Section~\ref{sec:UV}).  For these events low-energy X-ray observations are critical.  Although likely to be weaker than type Ib/c supernovae, it is possible to produce X-rays from interactions with the circumstellar medium.  We will discuss this in Section~\ref{sec:shockinteractions}.

\section{Type Ia Supernovae}
\label{sec:exIa}

Shock acceleration can also occur in the emergence of the supernova blastwave from its white dwarf progenitor in a thermonuclear supernovae.  Two past numerical studies of this emergence have argued that the emission will peak in the hard X-rays or even gamma-rays~\citep{2009ApJ...705..483H,2010ApJ...708..598P}.  Although both of these studies predicted high energy emission, they had very significant differences.  \cite{2009ApJ...705..483H} predicts photon energies of 100\,keV.  Their total emitted energy is roughly $10^{47}$\,erg of X-ray/gamma-ray energy with luminosities of above $10^{47} {\rm \, erg s^{-1}}$ for nearly half a second.  This is because they found that the shock accelerated dramatically as it emerged from the star in their pure hydrodynamic calculations.  This acceleration is limited to a narrow range that, in radiation hydrodynamics calculations will be smoothed over, leading to much lower peak velocities.  \cite{2010ApJ...708..598P} instead argues for roughly 20\,keV energies with a total emitted energy of $10^{40} {\rm \, erg}$ emitted in 0.01\,s.

We can directly apply our models to the SBO in type Ia, but with a slightly different prescription than our massive star models.  None of these models are extremely relativistic (typical Lorentz factors of the fastest ejecta are below 1.1), but at high densities ($0.01-100 {\rm \, g \, cm^{-3}}$), we would expect from Equation~\ref{eq:temp} that the temperature would reach values between 10--100\,keV, matching the results from these studies. However, at these high densities, the radiation should be trapped.  The mean free path of the ejecta is:
\begin{equation}
    \lambda_{\rm mfp} = (\rho \kappa)^{-1}
    \label{eq:mfp}
\end{equation}
where $\rho$ is the density and $\kappa$ is the opacity ($0.2~{\rm cm^2 g^{-1}}$ for electron scattering).  Although most of the radiation is trapped, if we assume that there is a sharp boundary between this ejecta and its surroundings are optically thin, this high energy emission should escape.  Assuming electron scattering and a density of $0.01 {\rm \, g \, cm^{-3}}$, the mean free path is 500\,cm.  If the shock is at $10^4$\,km, the amount of mass in this region would be $6 \times 10^{21} {\, \rm g}$ and the total energy in this region would be $6 \times 10^{41} {\, \rm erg}$.  As the ejecta expands, it adiabatically cools, so we expect only the initial breakout (across the first few mean free paths) to produce be high-energy emission.

To study SBO in type Ia, we assume peak velocities corresponding to a Lorentz factor of 1.1 with a steep Lorentz factor distribution:  $p=6$.  We then vary the breakout density which is dictated by the sharpness of the boundary between the exploding white dwarf and its circumbinary properties.  Depending on this progenitor, the circumbinary conditions can vary dramatically.  We consider three very different densities for the edge of the system at SBO:  $10^{-6}$, $10^{-2}$, $10^{2} \, {\rm g \, cm^{-3}}$.  The lower densities correspond to progenitors whose circumbinary density is higher because the ejecta must expand further to breakout.  These different densities correspond to three different initial shock-heated temperatures for the ejecta (from equation~\ref{eq:temp}): $\sim 1, 10, 100\,{\rm keV}$. This shock-heated ejecta expands adiabatically before the radiation can escape and power the observed signal.

With each density, we also calculate the mass within a few mean free paths of the ejecta edge, and hence, thermal energy of the ejecta emitting high-energy photons.  We assume that the total energy available to power our light-curve is limited to the ejecta within a few mean free paths of the edge of the star.  The mean free paths for our three densities are: 50\,km, 500\,cm, and 0.05\,cm.  As such, we expect much more ``emitting mass'' (and energy) if the density at SBO is lower. 

\begin{figure}
    \includegraphics[width=0.5\textwidth]{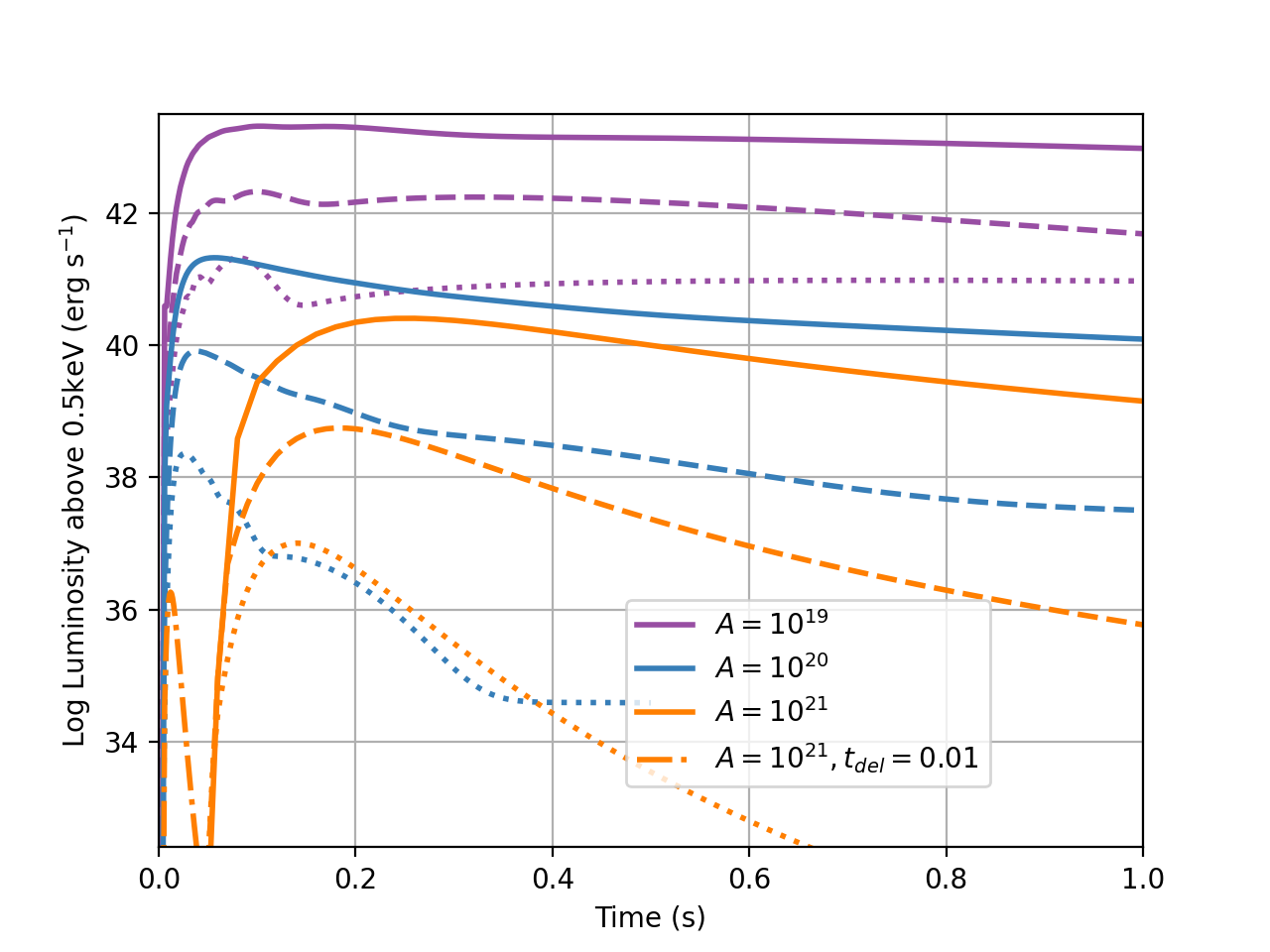}
    \caption{X-ray luminosities for a set of Type Ia SNe with 3 different expansion radii/areas ($10^{19}, \, 10^{20}, \, 10^{21} \, {\rm cm^2}$) and 3 different densities:  $10^{-6}$ (solid), $10^{-2}$ (dashed), $10^{2}$ (dotted) ${\rm g \, cm^{-3}}$.  These models behave differently than our core-collapse SN models as we are tying the emitting area and densities to the total energy with velocities of $\beta \Gamma>0.1$.  Although much less important than type Ia SNe, he escape time can play a role in core-collapse supernovae:  see Section~\ref{sec:other}.} 
    \label{fig:flxray_Ia_lum}
\end{figure}

Figure~\ref{fig:flxray_Ia_lum} shows results from our type Ia model simulations.  Our luminosities are much closer to the analysis in \cite{2010ApJ...708..598P}, but showcase the broad diversity depending on the nature of the edge of the star at the SBO which, in turn, depends on the circumbinary densities produced in the type Ia progenitor.  This SBO signal can then be used to probe these densities and, ultimately, the progenitor behind the explosion.

\begin{figure}
    \includegraphics[width=0.48\textwidth]{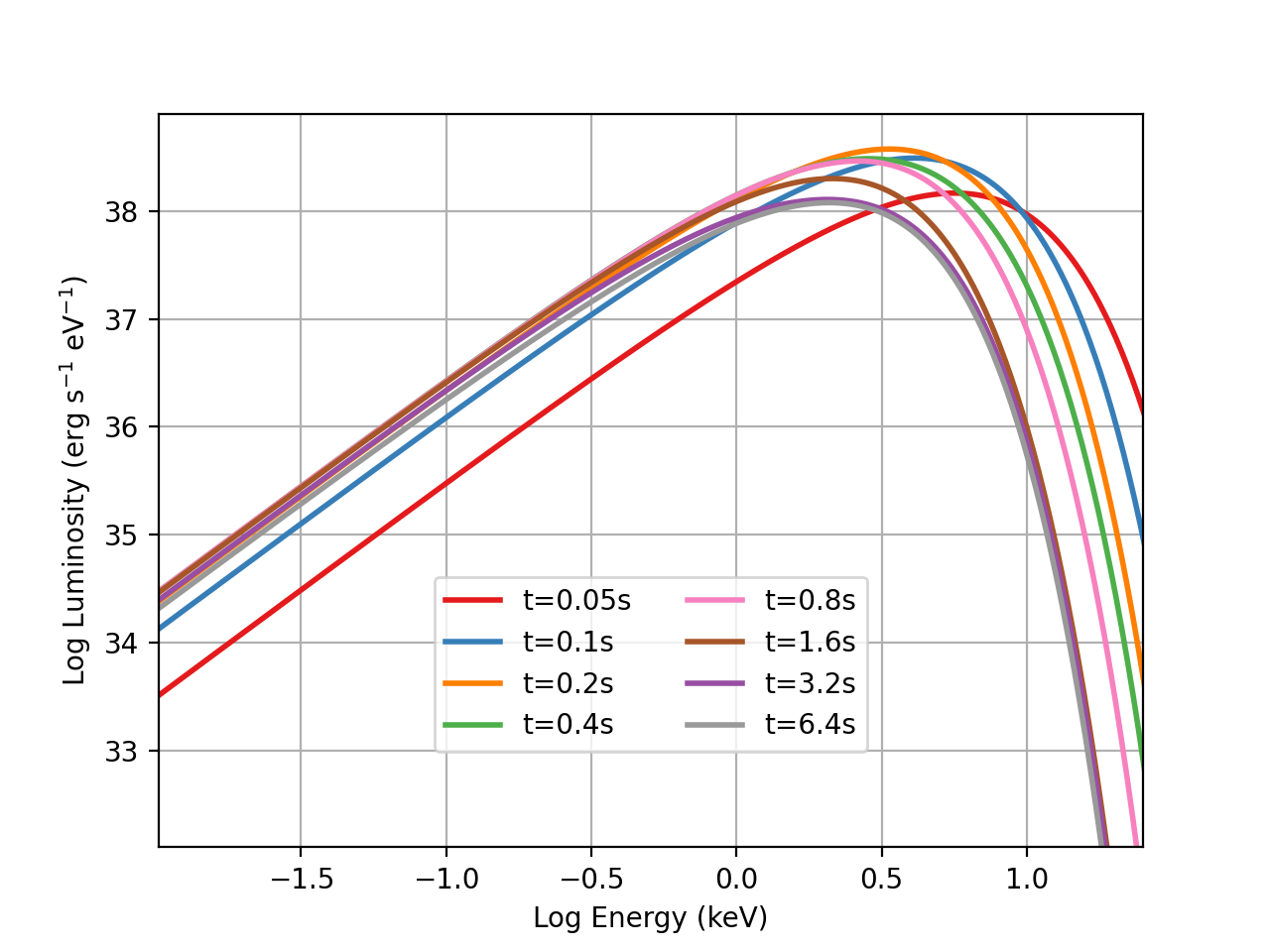}
    \includegraphics[width=0.48\textwidth]{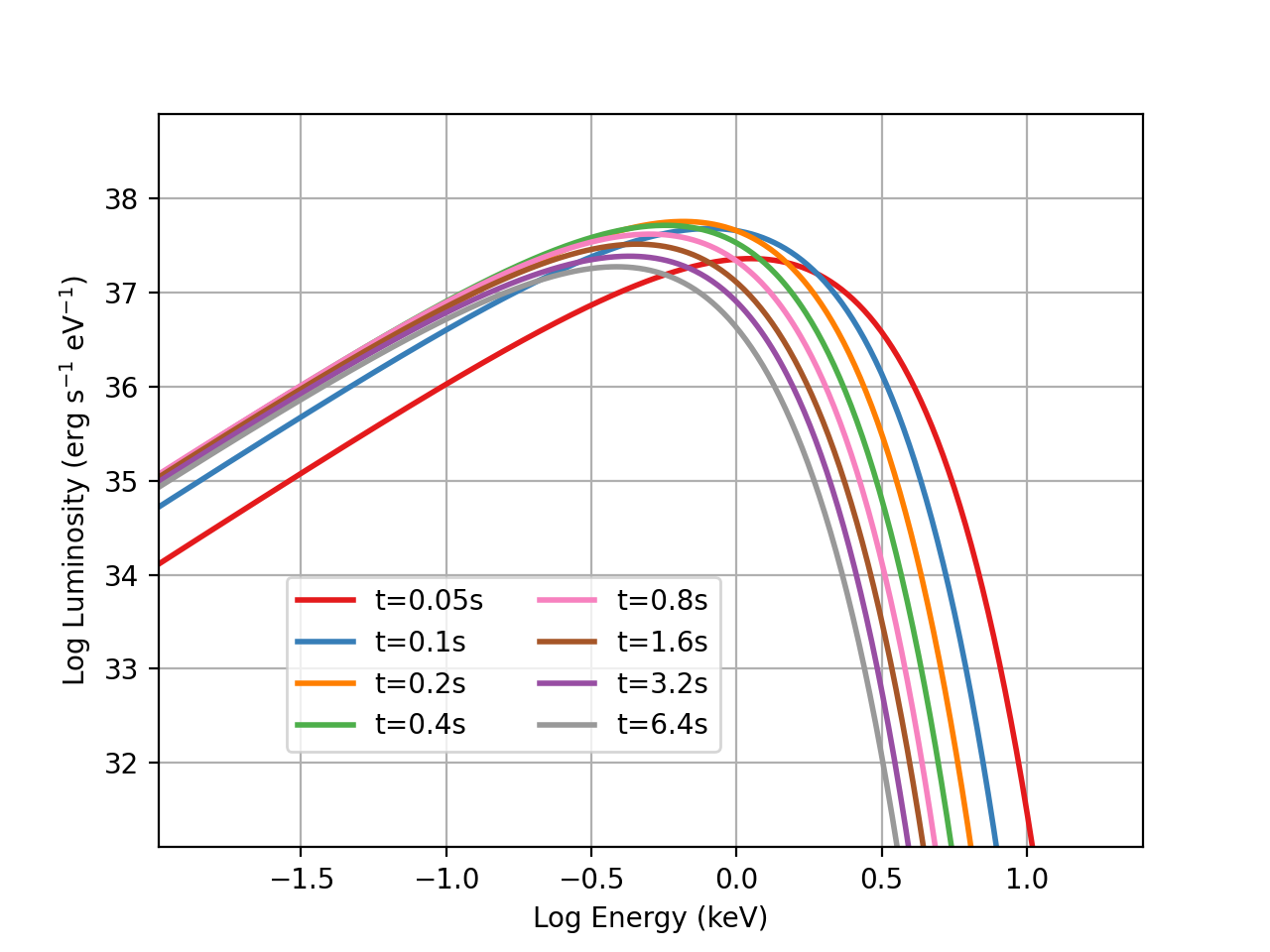}
    \caption{Spectra for two $10^{-6} {\rm g \, cm^{-3}}$ models from Figure~\ref{fig:flxray_Ia_lum} corresponding to expansion areas of $10^{19}$ (top) and $10^{20}$ (bottom) cm$^2$.  Ia emission peaks below 2-5\,keV and drops below 1\,keV in 1--10\,s.} 
    \label{fig:flxray_Ia}
\end{figure}

Figure~\ref{fig:flxray_Ia} shows the corresponding spectra for two of our models (expansion areas of $10^{19}, \, 10^{20} {\rm cm^2}$) with a density of $10^{-6}~{\rm g \, cm^{-3}}$ (relatively bright models).  Our Ia models peak at photon energies below 2-5\,keV, respectively, for the two models and within a second, the peak emission drops a factor of 3-5.  

\section{Other Effects}
\label{sec:other}

\subsection{Other Shock Interactions}
\label{sec:shockinteractions}

Other shock interactions can contribute to detected signals in these events, especially if these shock interactions occur at early times, enabling more detections but complicating physical inference.  These interactions occur when the blastwave interacts with material in the circumstellar medium, including clumpy wind material~\citep{2020ApJ...898..123F}, shells from binary mass-loss events~\citep{2017hsn..book..403S} and binary companions themselves~\citep{2000ApJS..128..615M,2010ApJ...708.1025K}.  The complete study of these interactions is beyond the scope of this paper, but the basic physics behind the thermal emission is the same.  Here we present some preliminary studies to compare these results to those of SBO and better determine the difficulties in disentangling these effects.

For this study, we assume a range of densities for the shock interactions.  As we discussed in Section~\ref{sec:SBOprop}, the total energy in the shock region depends both on the density of the interaction conditions (densities, density gradient, interaction area) and blastwave properties (velocity and density).  Our models span the different interaction possibilities:

{\bf Clumpy medium:} The densities of the clumps depend on the level of instabilities in the wind.  As the average wind density varies with radius, it will also vary with the wind location.  For strong Wolf-Rayet winds of Ib/c supernovae, the densities at the stellar surface can be as high as $10^{-13} {\rm \, g \, cm^{-3}}$, decreasing as $r^2$ (recall, $\dot{M}_{\rm wind} = 4 \pi r^2 \rho_{\rm wind} v_{\rm wind}$).  We consider clumps that are 100-1000 times more dense than the average wind density, focusing on clumps with densities of $10^{-11}, 10^{-13}$ and $10^{-15} {\rm \, g \, cm^{-3}}$.  For type II supernovae, the wind mass loss is lower and the radius is higher and, for these clumps, we include models with densities of $10^{-17}, 10^{-19} {\rm \, g \, cm^{-3}}$.

{\bf Shell Interactions:} Shells from explosive burning or common-envelope mass ejection can produce higher densities.  To affect the early-time emission, these shells must be nearby so we focus on shells that are within $10^{14}-10^{16} {\rm cm}$.  The density is a function of this distance, the mass of the shell and the thickness of the shell.  For our models, we consider shells that have $7 M_\odot$ of ejecta (mimicking a common envelope event) and a thickness of 10\% of the radius with densities  $10^{-8}, 10^{-11}, 10^{-14} {\rm \, g \, cm^{-3}}$.

{\bf Companion Interactions:}  A large fraction of core-collapse and thermonuclear supernovae occur in binary star systems.  When the supernova blastwave interacts with this companion, it heats the companion and causes it to expand.  In some cases, e.g. the potential progenitors of broad-line Ic supernovae~\citep{2025ApJ...986..185F}, the companion can be quite close to the exploding star (within $10^{10}-10^{11}$\,cm).  Shock densities can be quite high ($10^{-6}-1 {\rm \, g \, cm^{-3}}$).  This shock blastwave material will adiabatically expand until the radiation can escape to produce a transient.

\begin{figure}
    \includegraphics[width=0.5\textwidth]{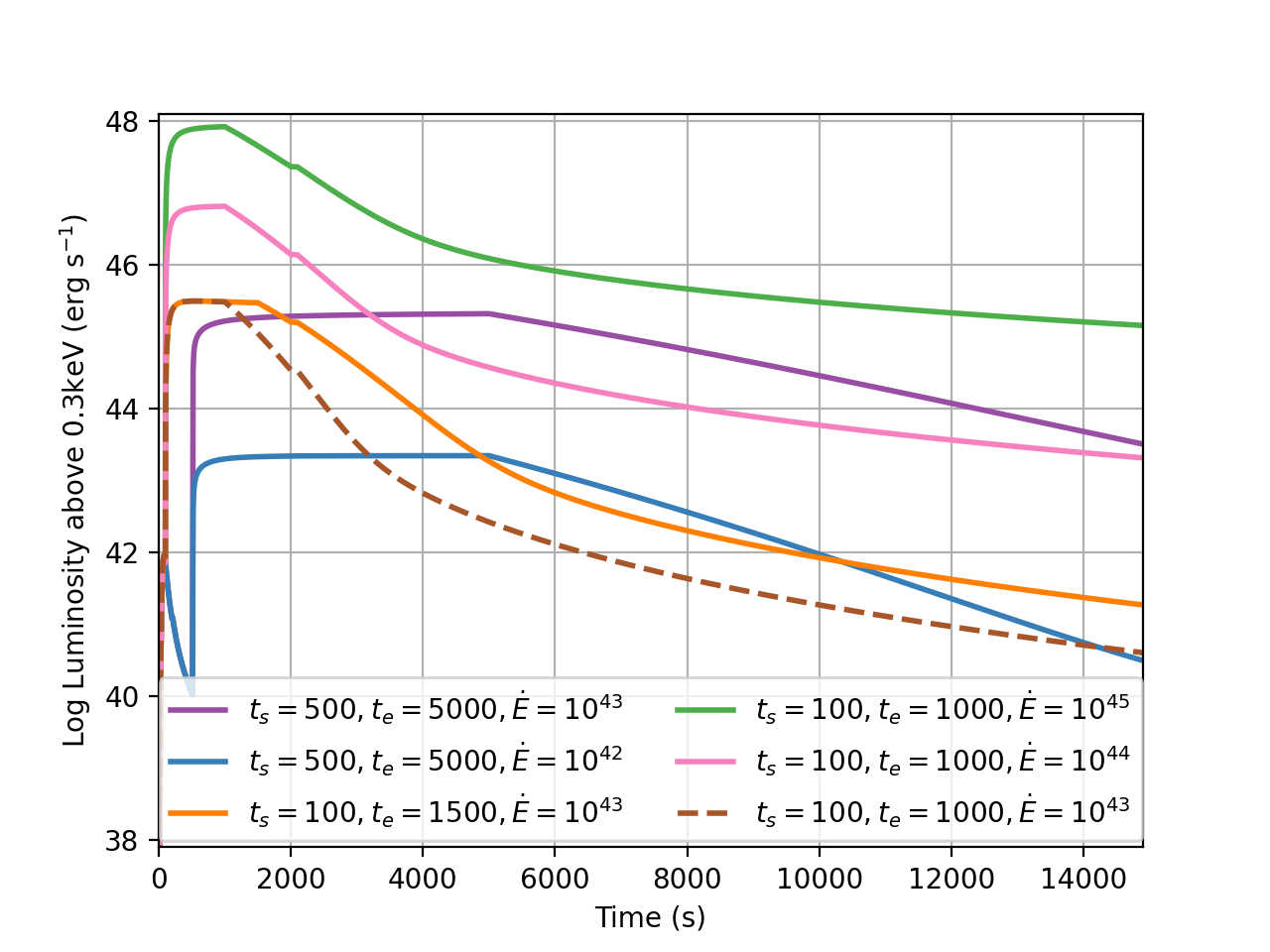}
    \caption{X-ray luminosities for a series of convective-engine Ic models ($p=6, \beta \Gamma_{\rm max} = 1.2$).  We have implemented a simplified shock heating model that adds energy ($\dot{E} dt$) to the system between start and end times of the shock heating ($t_s, t_e$).  The energy is distributed following the initial $\beta \Gamma$ distribution ($p=6$).  The initial models would normally produce a weak SBO signal, but the shock interactions revive the emission, producing strong X-ray signals.  Radiation-hydrodynamic models could help determine the amount of heating possible for a given supernova explosion.} 
    \label{fig:flxray_shocks}
\end{figure}

For all of these shock interactions, the additional heating will alter the evolution of the SBO cooling.  A systematic study of this physics is beyond the scope of this paper.  However, Figure~\ref{fig:flxray_shocks} shows a handful of examples of the effect shock heating can have on the light-curve.  Here we have artificially included an energy source from shock interactions.  More detailed shock heating have been implemented for SN light-curves, e.g. ~\cite{2025arXiv250115702N}.  For these models, we implement a simple model that injects energy using our initial velocity distribution.  For these models, we use $p=6, \beta \Gamma_{\rm max} = 1.2$.  We add energy ($\dot{E}_{\rm heat} dt$) to the system between start and end times of the shock heating ($t_s, t_e$). Typically, these models would produce a weak SBO signal.  But the additional energy added can drastically alter the X-ray signal.  Understanding these shock interactions, particularly just beyond the photosphere of the star/stellar wind, is critical to understanding the X-ray emission.

\subsection{Rise Time}
\label{sec:rise}

Another factor that affects the SBO signal is the initial rise time.  If we assume the shock-heating is limited to the outermost layer, the rise time is very rapid and can provide insight into the stellar radius.  But for the more complex shocks studied in this paper, the rise time also depends on the density of the ejecta and the extent of the shock-heated (``emitting'') material.  If the emitting material is distributed across a distance $\delta r$, its emission ($L$) is attenuated by roughly:
\begin{equation}
    L \sim \int_{r_{\rm ej}-\delta r}^{r_{\rm ej}} \epsilon r^2 dr e^{-(r_{\rm ej} - r) \rho \kappa} 
\end{equation}
where $\epsilon$ is the emissivity of the ejecta in ${\rm erg \, s^{-1} g^{-1}}$, $r_{\rm ej}$ is the radius of the ejecta, $\delta r$ is the extent of the emitting region, $\kappa$ is the opacity and $\rho$ is the density of the ejecta.  For many models (particularly explosions of compact stars), the density is sufficiently high that most of energy in the emitting region is trapped.  As the material moves outward, the radius and radial extent increases and the density decreases (roughly as $r_{\rm ej}^{-2} \delta r^{-1}$).  This layer will ultimately become optically thin, $e^{-\delta r \rho \kappa} \rightarrow 1$, and the shock will cool as it emits.  

\begin{figure}
    \includegraphics[width=0.5\textwidth]{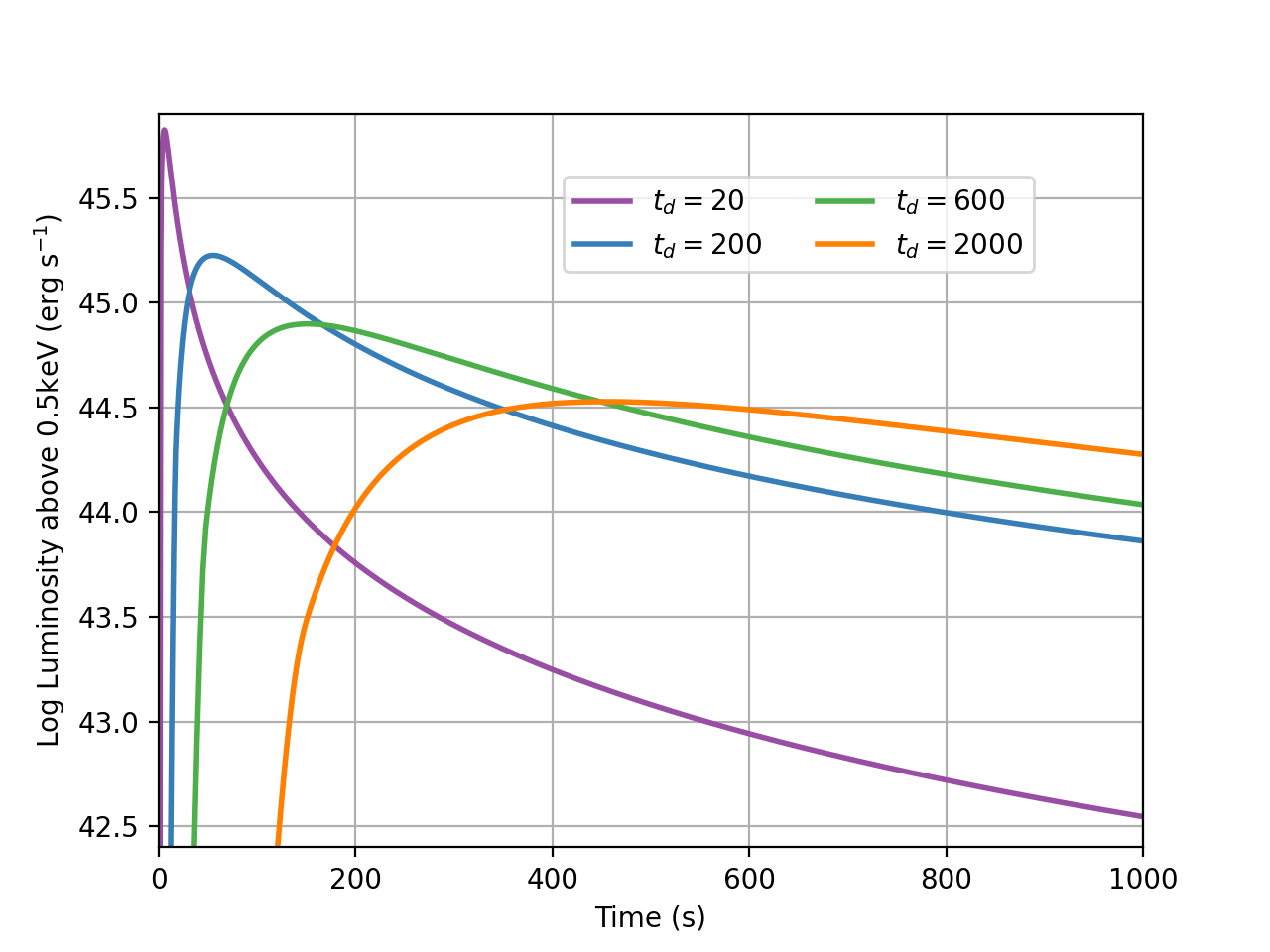}
    \caption{The X-ray luminosity from a strong convective-engine explosion (power law distribution $p=4$, total energy with $\beta \Gamma > 0.1$ set to $10^{50}\,{\rm erg}$, emitting area set to $10^{24} {\rm \, cm^2}$) varying the emergence time ($t_d$) from 20-2000s.  The Shorter emergence times peak sooner and higher, but cool the ejecta faster, leading to lower late-time luminosities.} 
    \label{fig:tdelay}
\end{figure}

We mimic this escape process by instigating a simple modifier to the emission $e^{-t_{d}/time}$.  This modifier not only affects when the emission peaks, but also how quickly the shock cools.  Figure~\ref{fig:tdelay} shows the X-ray luminosity above 0.5\,keV for a strong convective engine model where we have varied the emergence time from 20 to 2000\,s.  Shorter emergence times peak sooner and higher, but cool the ejecta faster, leading to lower late-time luminosities.  By combining measurements of the emergence time with velocity estimates and total emission, we can place constraints on the density of the ejecta and, hence, the stellar edge.  We can also constrain the extent of the shock-heated region.

\subsection{Ultraviolet Emission}
\label{sec:UV}

\begin{figure}
    \includegraphics[width=0.5\textwidth]{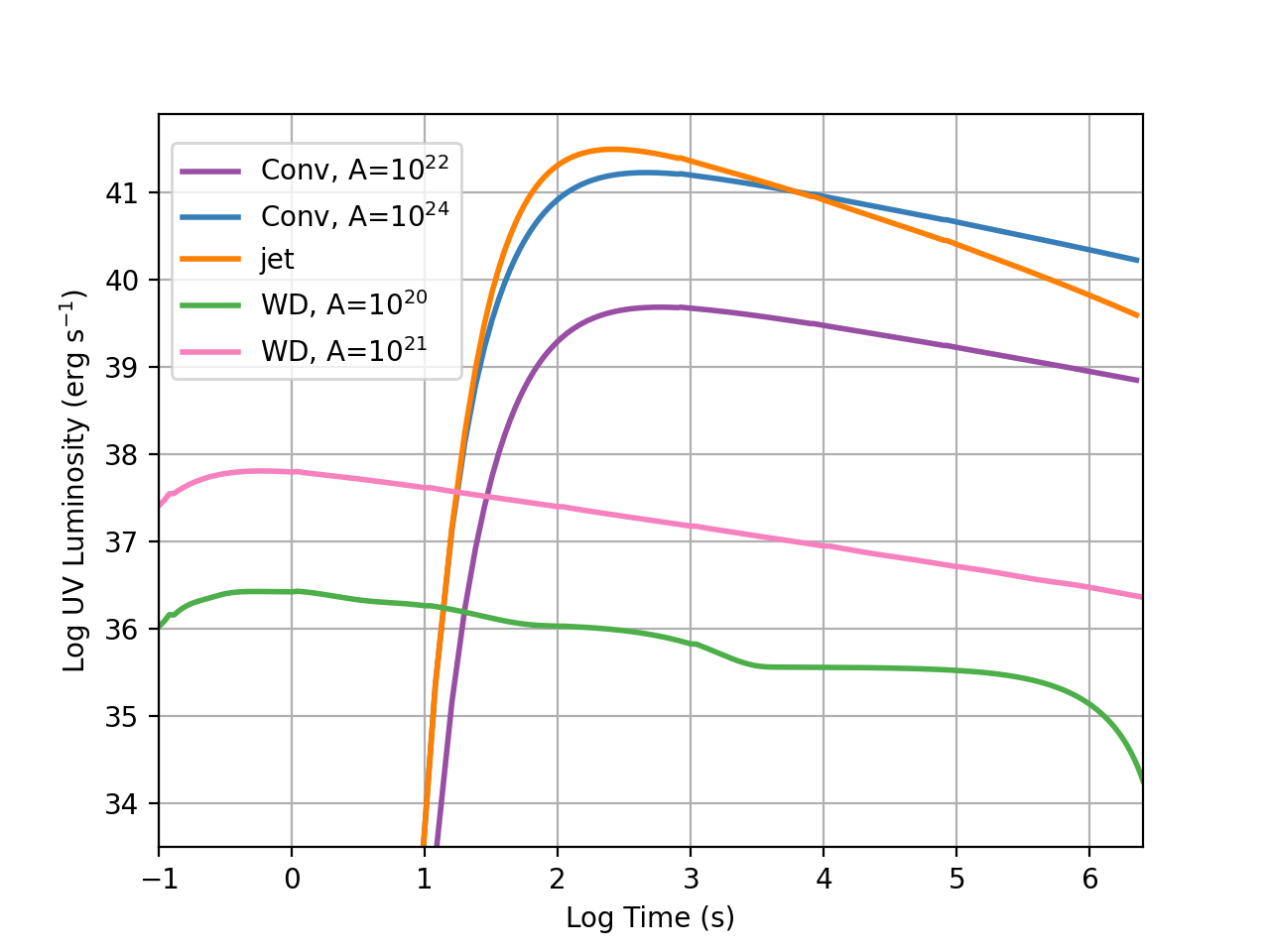}
    \caption{UV light-curves for a subset of our models.  Included in this set are two convective-engine models:  $\Gamma_{\rm max} = 10$, $p=4$ with a total energy of $10^{48} \, {\rm erg}$ with $\beta \Gamma > 0.1$ and two emitting areas ($10^{22}, \, 10^{24} {\rm \, cm^2}$).  We also have one jet-driven model with $\Gamma_{\rm max} = 50$, $p=1$ with a total energy of $10^{48} \, {\rm erg}$ with $\beta \Gamma > 0.1$ and an emitting area of $10^{24} {\rm \, cm^2}$.  Finally, we include two type Ia supernova models with two expansion radii/areas ($10^{20}, \, 10^{21} \, {\rm cm^2}$) for the $10^{-2}$ density.}
    \label{fig:fuv_fin}
\end{figure}

In many cases, the peak of the SBO and cooling emission occurs below photon energies of 0.1\,keV (Figures~\ref{fig:stripped_spec}, \ref{fig:specII}).  Although photon energy of this peak emission is above ultraviolet bands, UV observations can provide complementary information to help differentiate different explosion models.  We defer a thorough study of the complementary nature of UV and X-ray observations to a future paper.  But to get a rough feeling of the luminosities and durations of the UV signals from our models, we show a few UV light-curves (emission between 100 and 250\,nm) from a select set of our models in Figure~\ref{fig:fuv_fin}.  The UV emission tends to peak much later than the X-ray. However, it is important to note that, at least with these models, the peak luminosities are much lower than the X-ray emission.  

\subsection{X-ray Propagation Effects}
Unfortunately, extinction and absorption can limit observations of SBO from far ultraviolet to soft X-rays. This must be considered in rates detections for future missions and interpretation of spectral observations of specific events. The cumulative extinction is typically modeled as an integrated hydrogen column density, $N_H$, with contributions from the Milky Way and the host galaxy. Predicted extinction values as a function of source position exist for the Milky Way, but these may underestimate the total effect \citep{willingale2013calibration}. Host extinction values vary greatly, and there are typically trends for different extragalactic source classes (e.g. due to different preferred host types, preferred galaxy positions, or preferred metallicities, etc). There are only a few viable measurements of host galaxy $N_H$ for (candidate) SBO events, with values of $(4-10)\times 10^{21}$~cm$^{-2}$ for a Ib and three Ic-BL events, as detailed below. Galactic contributions are typically below these values. These can begin to have effects at $\sim$2~keV, and $\sim$50-80\% losses at $\sim$1~keV. Extreme extinction can affect up to $\sim$5~keV, while low extinction can matter below $\sim$0.2~keV.

While extinction has a large range and observations are performed on individual events, we briefly comment on how we expect it to affect supernovae by type, guided by the small observed sample. For type II supernovae the peak emission likely occurs in an inaccessible wavelength range, limiting observations to ultraviolet emission as the initial shock expands and cools. For our models, Ic-BL SBO is generally hard enough that extinction effects are largely negligible. Type Ia and convective engine Ibc have peaks at energies where extinction/absorption can have important effects.


\section{Archival Events}
\label{sec:obscomp}

Temporal evolution, time-integrated spectral results, and time-resolved spectral analysis are powerful observables for using SBO to learn about both the explosion and stellar progenitor. In order to validate these models, and identify obvious limitations, we compare with archival transients which have been interpreted as high-energy detections of SBO events. To summarize the findings of these sections we find that 
\begin{itemize}
    \item the XRF~080109 seen before the type Ib SN~2008D is consistent with a pure thermal SBO origin
    \item GRBs like 060218 and 100316D, which are X-ray rich and extremely long low luminosity GRBs followed by Ic-BL supernovae, are inconsistent with a pure SBO origin, but are viable for cases with shock interactions with immediately nearby and dense circumstellar medium, so long as the velocity gradient is shallow.
    \item Other low luminosity GRBs like GRB~980425, followed by the Ic-BL SN~1998bw, are inconsistent with a SBO origin which considers only thermal bremsstrahlung (though may be viable with an additional synchrotron component, without a jet \citep{2001ApJ...551..946T})
    \item The low energy excess identified in prompt GRB spectra, which are sometimes considered to be thermal from photospheric (SBO) emission, may struggle with the $\sim$1~s cooling timescales which result from high $\Gamma$ values.
    \item Many Einstein Probe extragalactic transients are consistent with a SBO origin.
\end{itemize} 
\noindent There are, unfortunately, no compelling examples of SBO in the X-ray from type II or Ia events.

\subsection{XRF 080109 / SN 2008D}
The serendipitous detection of XRF~080109 by the Swift-XRT \citep{burrows2005swift} preceded the associated type Ib SN~2008D and is a clear SBO signature \citep{2008Natur.453..469S,mazzali2008metamorphosis,Malesani_2009}. The XRT lightcurve, covering 0.3-10~keV, shows a $\sim$70~s rise time with a $\sim$500~s decline. The peak luminosity is $L_x \approx 6\times10^{43}$~erg~s$^{-1}$ and total energy is $E_x \approx 2 \times 10^{46}$~erg. \citet{2008Natur.453..469S} perform time-integrated fits which show a preference for a power-law over a blackbody, which they suggest implies any thermal component is under 0.1~keV and attribute the signature to non-thermal emission from bulk Comptonization \citep{2007ApJ...664.1026W}. The power-law fit gives an index of $2.3 \pm 0.3$. As we show below, a power-law index from thermal Bremsstrahlung is consistent with expectations, once temporal evolution is considered.



\begin{figure}
    \includegraphics[width=0.48\textwidth]{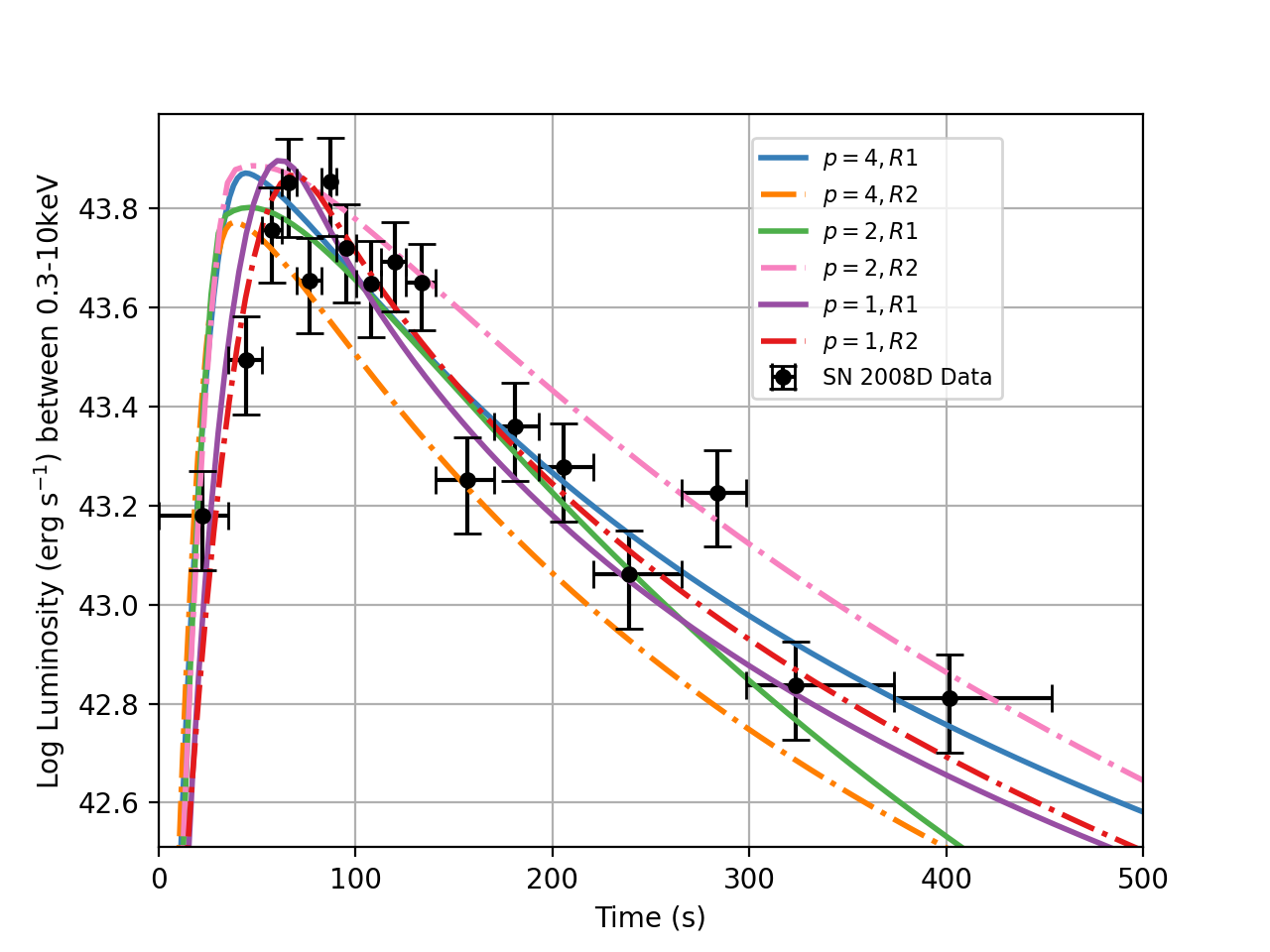}
    \includegraphics[width=0.48\textwidth]{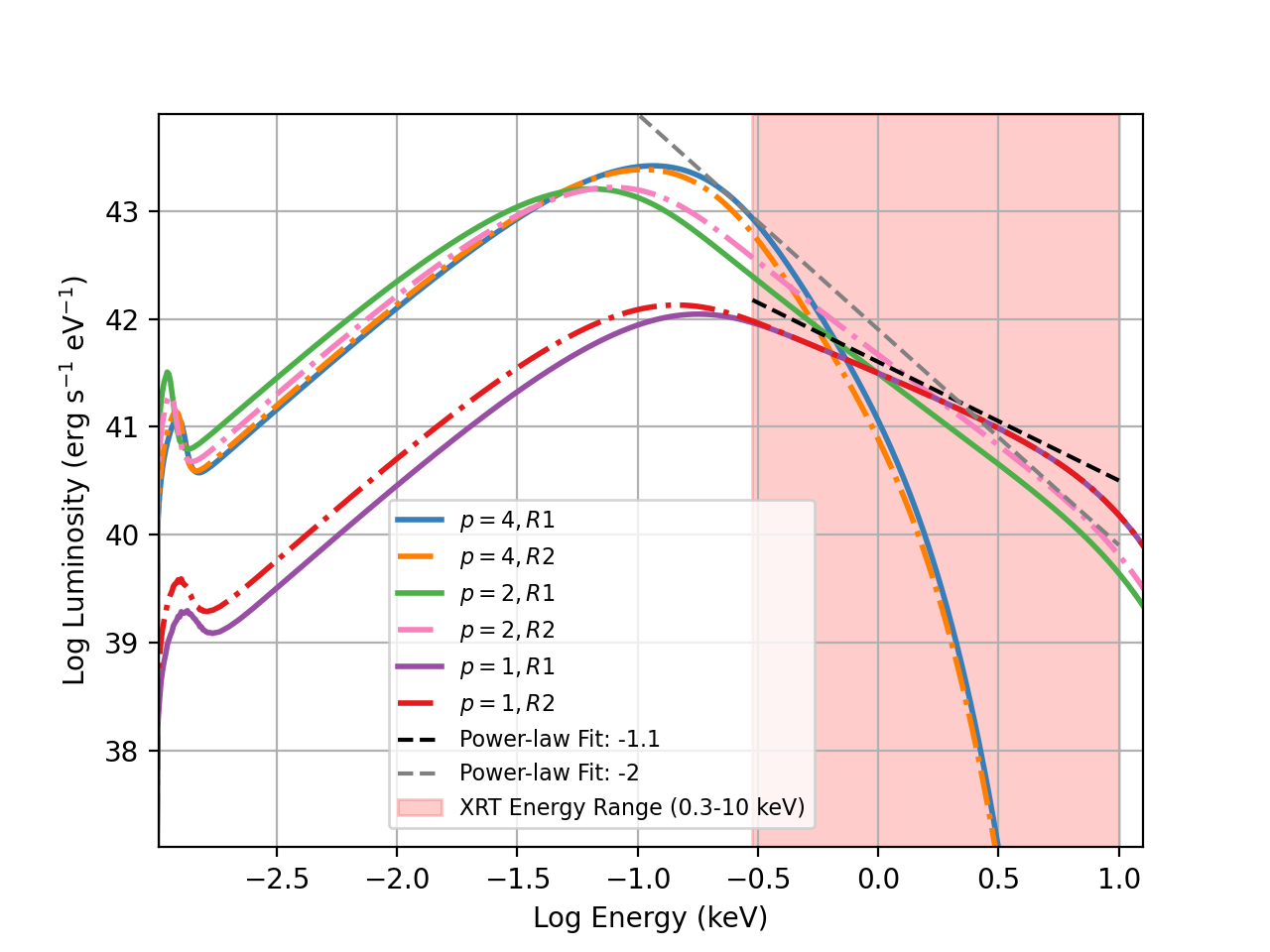}
    \caption{Light-curves and integrated spectra for a set of runs designed to match SN2008d.  The $p=4$ runs $R1,R2$ have, respectively, emitting areas of $0.8,1.0 \times 10^{23} \, {\rm cm^2}$ and a total energy of $1.8,2.7 \times 10^{48} \, {\rm erg}$ above $\beta \Gamma>0.1$, $R2$.  The $p=2$ runs $R1,R2$ have emitting areas of $5, 3 \times 10^{22} \, {\rm cm^2}$ respectively, both with a total energy of $6 \times 10^{46} \, {\rm erg}$ above $\beta \Gamma>0.1$, $R2$.  The $p=1$ runs $R1,R2$ have, respectively, emitting areas of $1,3 \times 10^{20} \, {\rm cm^2}$ and emergence times of 300, 500\,s.  Both have a total energy of $2 \times 10^{46} \, {\rm erg}$ above $\beta \Gamma>0.1$, $R2$.}
    \label{fig:modelsn2008d}
\end{figure}

Figure~\ref{fig:modelsn2008d} shows a range of models against the observations of SN~2008D, finding a self consistent SBO picture. These models use 3 different values for the power ($p$) of the Lorentz-factor distribution, each with two different runs varying the total energy at the relativistic ejecta.  The $p=4$ runs $R1,R2$ have, respectively, emitting areas of $0.8,1.0 \times 10^{23} \, {\rm cm^2}$ and a total energy of $1.8,2.7 \times 10^{48} \, {\rm erg}$ above $\beta \Gamma>0.1$.  The $p=2$ runs $R1,R2$ have emitting areas of $5, 3 \times 10^{22} \, {\rm cm^2}$ respectively, both with a total energy of $6 \times 10^{46} \, {\rm erg}$ above $\beta \Gamma>0.1$.  The $p=1$ runs $R1,R2$ have, respectively, emitting areas of $1,3 \times 10^{20} \, {\rm cm^2}$ and emergence times of 300, 500\,s.  Both have a total energy of $2 \times 10^{46} \, {\rm erg}$ above $\beta \Gamma>0.1$. The smooth rise and decay are compatible with the X-ray signal arising from SBO, and we thus do not explore further contributions (e.g. circumstellar medium interactions). A number of models are compatible with the temporal profile, though we note the rise time is longer than our base models, consistent with expectations. We additionally show the time-integrated spectrum output over a range of models, showing compatibility power-law values. Note the -2.3 photon index corresponds to a -1.3 index in energy space. This spectrum excludes models which are viable when comparing only with temporal information. A further exploration of the full breadth of parameter space allowed by the XRT observations of SN~2008D, including time-resolved spectral evolution comparison, is beyond the scope of this paper.


\subsection{GRBs 060218, 100316D, and Similar}
GRBs~060218 and 100316D belong to a unique class of low luminosity GRBs. GRB~060218 has a Swift-BAT \citep{barthelmy2005burst} duration over 15-150~keV of $\gtrsim$2~ks, a flat XRT temporal evolution peaking at $\sim$1~ks and lasting $\sim$10~ks, and a slow rise in ultraviolet and optical (UVOT) \citep{roming2005swift} consistent with shock emergence, followed by a Ic-BL supernova \citep{campana2006association}. The bolometric energy release is about $\sim6\times10^{49}$~erg with a peak luminosity of in gamma-rays of $\sim5\times10^{46}$ (and somewhat lower in X-rays). The time-integrated $E_{\rm peak} \approx 5$~keV. \citet{margutti2013signature} infer a shallow $\gamma \beta$ power-law index from the supernova to high energy ejecta. GRB~100316D appears generally similar, though with less complete information \citep{starling2011discovery}. Here we present a series of models to better infer the properties of this transient.

\begin{figure}
    \includegraphics[width=0.5\textwidth]{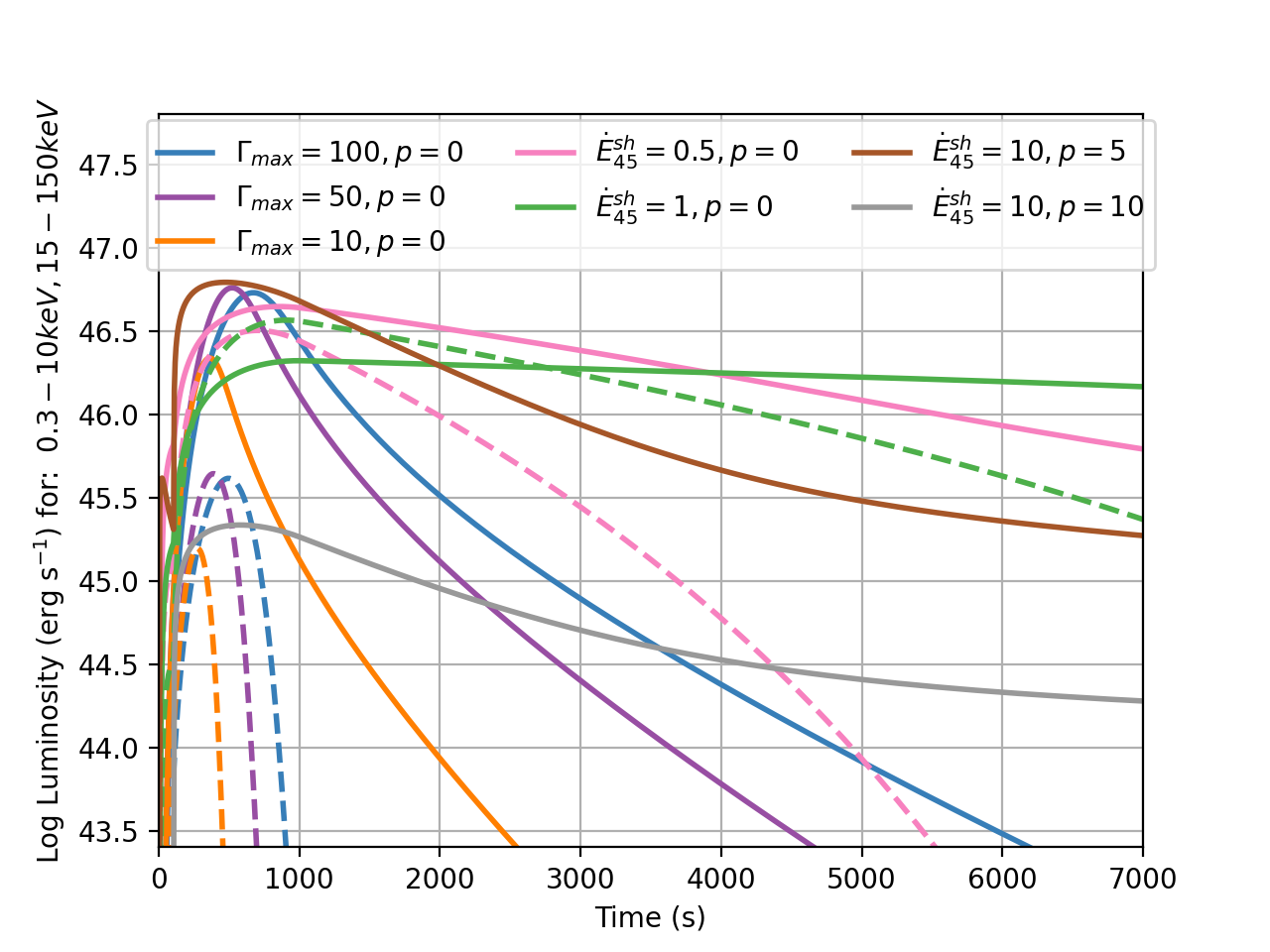}
    \caption{X-ray luminosity (0.3-10\,keV:  solid lines) and $\gamma$-ray (15-150\,keV:  dashed lines) emission for a range of Ib/c models designed to fit GRB~060218.  We modeled a series of models varying the emitting area, total energy at high velocities, and expansion velocities for a different maximum Lorentz factors:  $\Gamma_{\rm max} =10, 50, 100$.  None of these models match the flat X-ray curve nor could they explain more than $\sim 10\%$ of the $\gamma-$ray emission.  We then implemented a shock heating model ($\Gamma_{\rm max} =10$) where we deposited a small amount of the kinetic energy from 100-1000\,s.  These models can reproduce the flat X-ray signal and contribute much more significantly to the $\gamma-$ray (15-150\, keV) luminosity.}
    \label{fig:modelsn2006aj}
\end{figure}

Among the most unique properties of this burst is the long-lived X-ray emission over thousands of seconds. We tested a range of Lorentz factors for this model to determine the extent at which the BAT signal can be produced through Doppler-shifted Bremsstrahlung emission (Figure~\ref{fig:modelsn2006aj}).  We assume a jet-driven distribution of Lorentz factors ($p=0$) with maximum Lorentz factors ($\Gamma_{\rm max}$) ranging from 10-100.  Even with $\Gamma_{\rm max} =100$, our Bremsstrahlung emission model can only account for 10\% the total $\gamma$-ray emission.  These models also do not fit the flat X-ray light-curve.  However, our shock heating models were able to produce both increased $\gamma$-ray emission and extremely flat X-ray light-curves (the flat light-curve corresponds to the duration of the shock interaction as the blastwave propagates through the inhomogeneous circumstellar medium).  If we attribute the X-rays to Bremsstrahlung emission, we can place strong constraints on the early-time shock interactions (presumably from the blastwave propagating through a clumpy wind medium). 

\begin{figure}
    \includegraphics[width=0.5\textwidth]{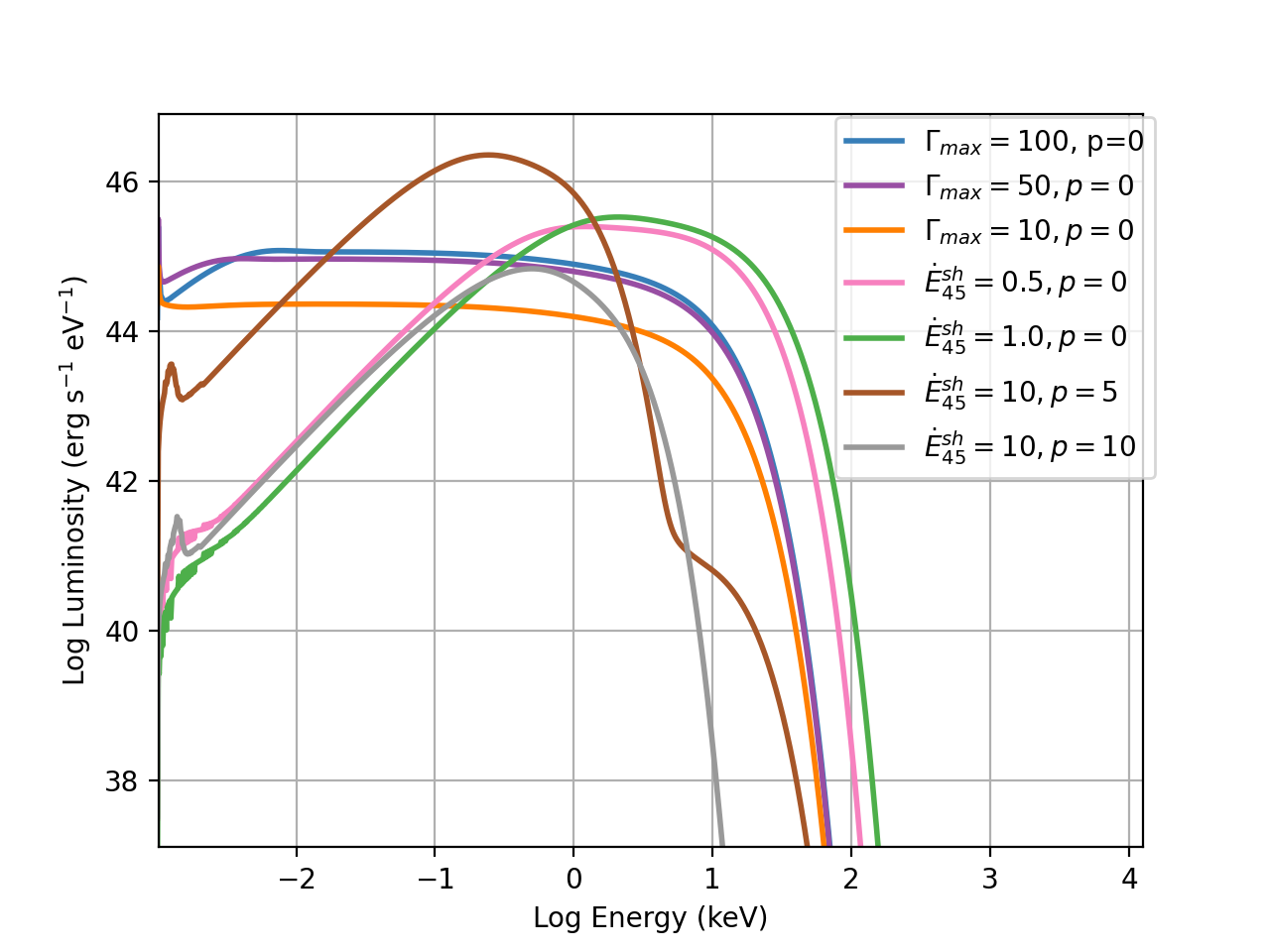}
    \caption{Time-integrated spectra for the models shown in  Figure~\ref{fig:modelsn2006aj}.  If we measure the integrated spectra but are limited to an energy range of 0.2-10keV (that of the Swift XRT), we could infer a range of power-law fits to these models that depend sensitively on the amount of high Lorentz factor ejecta.}
    \label{fig:modelsn2006ajspec}
\end{figure}

For most of our models, we assume $p=0$ for the distribution of Lorentz factors in the ejecta, arguing that the jet/cocoon model produces a large amount of ejecta at the higher Lorentz factors.  We included a couple shock heating simulations with much higher Lorentz-factor distributions: $p=5,10$.  These models effectively cut off the amount of high Lorentz factor ejecta.  The extreme case $p=10$ is only able to explain part of the observed X-ray emission and none of the gamma-ray emission.  Spectra can be used to probe this power law but the inferred velocity distributions will be different than those inferred assuming synchrotron emission.  Figure~\ref{fig:modelsn2006ajspec} shows the integrated spectra for our set of models.  The spectra are a summation of blackbody emitters and can take on a broad range of features.  Depending on the observations, these can be fit by a blackbody or a power law. We thus conclude that these events are consistent with shock breakout in a dense medium immediately surrounding the stellar progenitor, and with a shallow velocity gradient. This is self consistent with expectations of Ic-BL arising from jetted explosions.

\begin{figure}
    \includegraphics[width=0.5\textwidth]{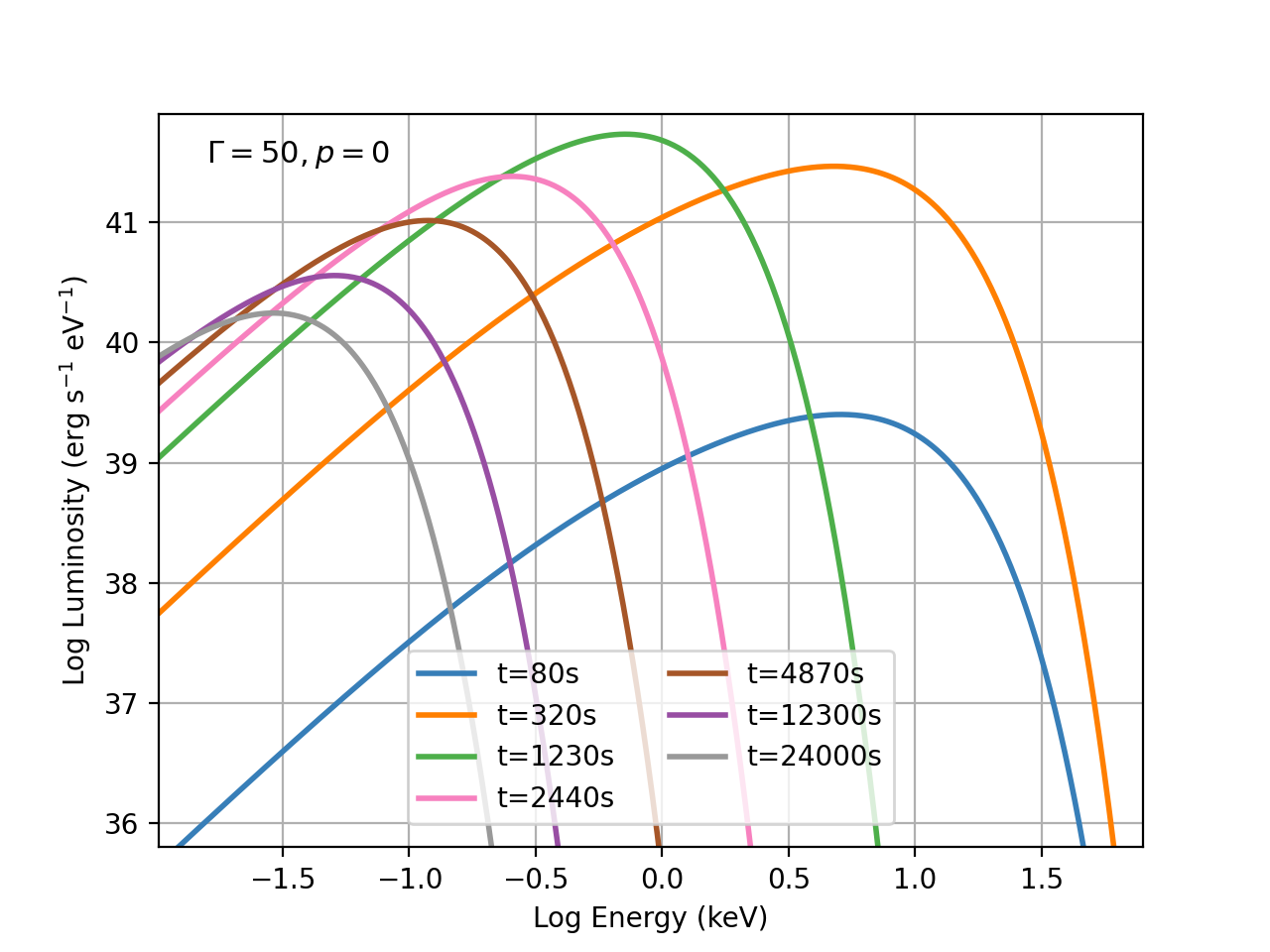}   \includegraphics[width=0.5\textwidth]{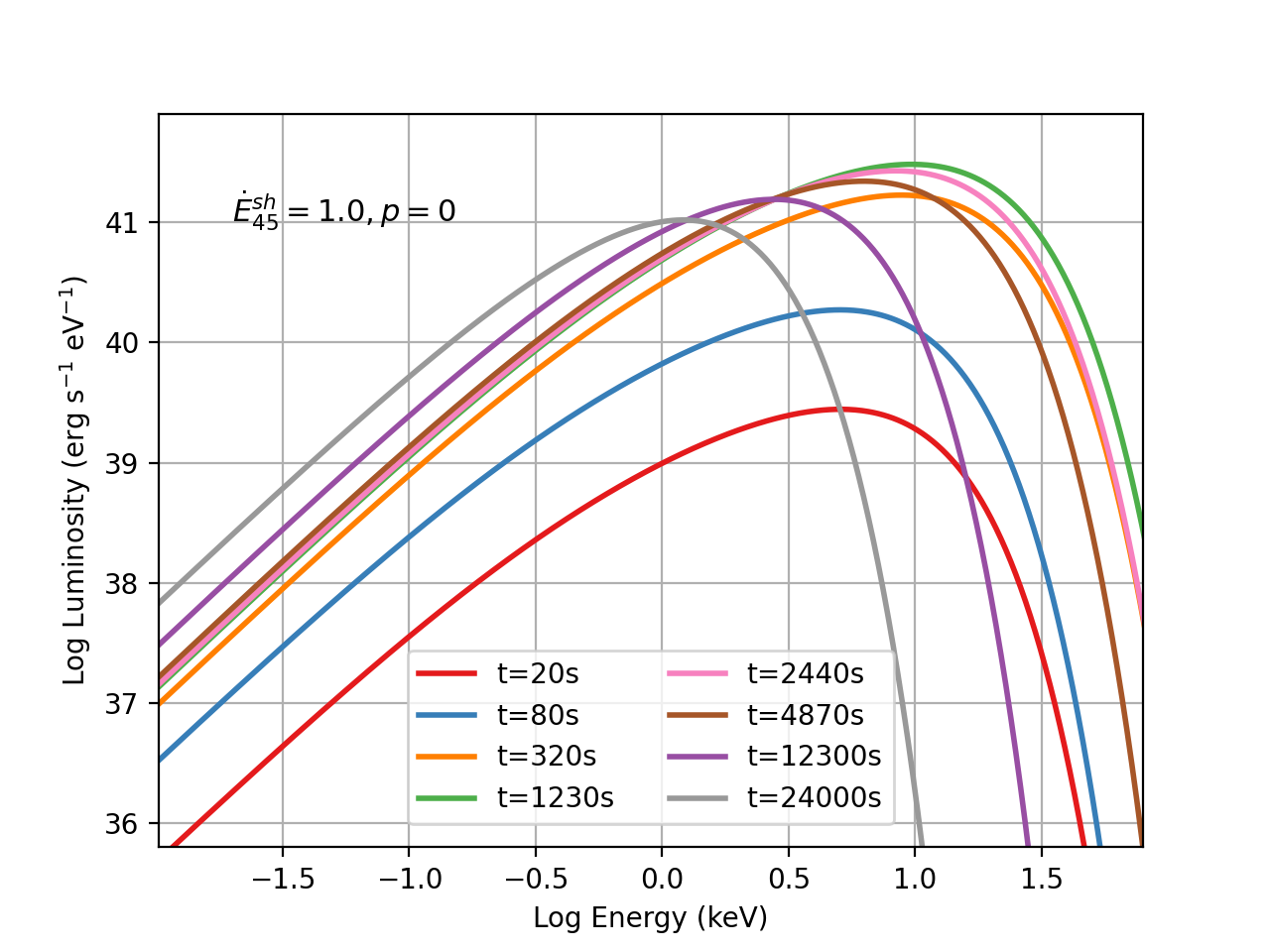}
    \caption{Spectra over a series of timesteps for two of our models shown in  Figure~\ref{fig:modelsn2006aj}.  The top panel shows the $\Gamma=50$ simple SBO model.  The bottom shows snapshots in time for one of our shock-heated models.  For simple SBO, the spectra is initially higher energy and slowly cools as it expands.  With shock-interaction models, the high-energy photons can be produced at later times.}
    \label{fig:sn2006ajspecevol}
\end{figure}

However, more could be learned with time-resolved spectral studies. Multiple studies investigate time-resolved spectral evolution with two component models, typically one being a blackbody \citep[e.g.][]{campana2006association,kaneko2007prompt,starling2011discovery}. The evolution of the spectra with time can be very different between simple SBO models and ones with late-time shock interactions.  Figure~\ref{fig:sn2006ajspecevol} shows the spectra for a range of times for two of our models for GRBs 060218 and 100316D.  Our fits use mild Lorentz factors that produce spectra that peak below 100\,keV.  In our normal SBO models, the high energy photons are produced at breakout and then the spectra reddens as it cools.  The shock-interaction models can reheat, producing high energy photons at later times.

\subsection{GRB 980425 / SN 1998bw}

The detection of SN~1998bw following GRB~980425 was the first direct proof that some GRBs arise from collapsars \citep{galama1998unusual,woosley2006supernova}. Even now it is the closest GRB detected and has the lowest luminosity of any GRB seen. The gamma-ray data from BATSE shows a prompt gamma-ray duration of $\sim$35~s, a time-integrated $E_{\rm peak}\approx140$~keV \citep{paciesas1999fourth,goldstein2013batse}. BeppoSAX reports both X-ray and gamma-ray data, showing the X-rays have a somewhat slower rise and longer duration, buts till ending in $\sim$50~s. The gamma-ray $E_{\rm iso}$ is $7\times10^{47}$~erg while the X-ray $E_{\rm iso}$ is only $\sim3\times10^{46}$~erg. The $3\sigma$ X-ray upper limits are $<10^{44}$~erg~s$^{-1}$ over $\sim$80-2000~s, proving a distinct behavior from GRBs~060218 and 1001316D.

Similar to those bursts, although the X-ray can be fit with thermal SBO models, these models predict a gamma-ray emission that is an order of magnitude below the observed emission.  We note that by including a simple synchrotron model, \cite{2001ApJ...551..946T} found that models similar to our simple SBO model could explain this GRB without a relativistic jet.  Thermal emission from shock interaction models with the right material densities can produce this gamma-ray emission.  But, with our formalism, we would require $\Gamma_{\rm max} >50-100$ and tuned material densities for the shock interactions.  The synchrotron model is probably more appropriate for the gamma-ray emission of GRB~980425, as well as other similar low luminosity GRBs. 


\subsection{Thermal Emission in Prompt GRBs}
The most extreme events of relevance are on-axis relativistic GRBs. These objects are incredibly energetic and involve the highest bulk Lorentz factors seen, allowing SBO emission to reach into the $\sim$MeV regime, classically referred to as photospheric emission. A major discovery of the Fermi-GBM \citep{meegan2009fermi} is an excess of emission around $\sim$tens of keV above the expectation from the broadly used empirical models. Many interpret this excess as arising from thermal photospheric emission \citep[e.g.][]{page2011grb,guiriec2011detection,guiriec2013evidence}, and even some claims that the full prompt emission of some bursts could be thermal \citep[e.g.][]{pe2012connection}. Further, there have been claims of thermal signatures in early X-ray follow-up observations \citep{page2011grb,starling2012search,sparre2012search}. Additionally, very nearby short GRBs show a soft tail extending beyond the typical short hard spike and is consistent with cooling thermal emission \citep{goldstein2017ordinary,burns2018fermi}. However, alternative explanations remain, e.g. the low energy excess could arise from an additional break in the prompt spectrum, as expected from fast-cooling synchrotron emission \citep[e.g.][]{ravasio2018consistency}. Thus, it is crucial to understand the viable range of thermal signatures expected in the prompt phase of these events. 

\begin{figure}
    \includegraphics[width=0.5\textwidth]{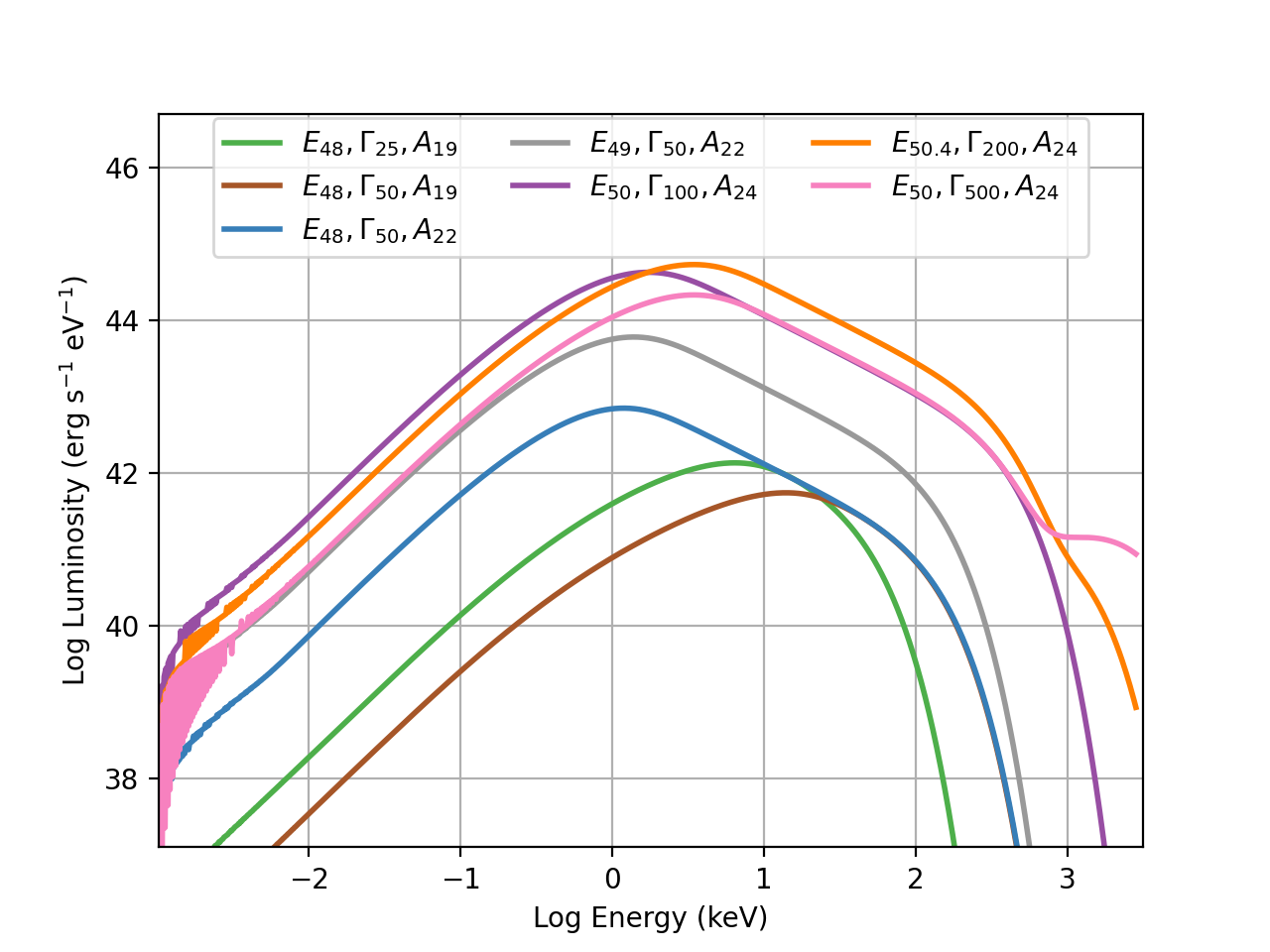}
    \caption{Time-integrated spectra for a range of jet-driven models including a subset of the models discussed in Section~\ref{sec:exjet} and some extreme jet models with Lorentz factors up to 500.  In all cases, we use $p=1$ for the Lorentz factor distribution.  But we vary the energy in the high velocity ejecta ($\beta \Gamma>0.5$), emitting area (from $10^{19}-10^{22} {\rm \, cm^2}$) and peak Lorentz factor ($\Gamma_{max}$).  For the most of these models, most of the photon energy is emitted between 1-100\,keV (peaking around 1-10\,keV.  But for higher Lorentz factors, a small amount of emission occurs above 1\,MeV.}
    \label{fig:GRBspec}
\end{figure}

The temporal evolution of the spectra for our strong, on-axis jet models, which have high Lorentz factors are shown in Figure~\ref{fig:GRBspecevol}. Although the photon energy can exceed 1\,MeV, the emission of these high-energy photons is fleeting. The peak energy drops below $\sim$30~keV in under half a second. Alternatively, one can look at time-integrated spectra from possible SBO emission from the jet, as shown in Figure~\ref{fig:GRBspec}. In this case, the spectral model would not be a basic blackbody, given the strong temporal evolution. Thus, naively, our model would expect thermal signatures in GRBs to rapidly evolve. However, this can be modified in cases where internal processes in the jet may give rise to multiple SBO events (e.g. internal shocks), which occurs with varying bulk Lorentz factors. However, each individual case should still have rapid evolution, providing an observational test.

\subsection{Einstein Probe Transients}
Einstein Probe is a new X-ray satellite which contains a sensitive, wide-field X-ray telescope \citep{2015arXiv150607735Y}. It is routinely detecting extragalactic X-ray transients. These appear to predominantly arise from gamma-ray bursts and Ic-BL supernovae \citep{o2025redshift}. However, the different energy range allows for the discovery of transients too soft to trigger gamma-ray burst monitors, or for characterization of additional components, including SBO. The most distant events generally require non-thermal processes, i.e. the X-ray signal is directly related to gamma-ray burst emission, so we focus on the closest events.

EP250108a is at a redshift of 0.176 with behavior strikingly similar to GRBs~060218 and 100316D, and is thus compatible with a SBO origin with shock heating contributions from nearby circumstellar material interactions \citep{eyles2025kangaroo,rastinejad2025ep,srinivasaragavan2025ep250108a}. EP240414a, EP250304a and EP250827b show similar consistency \citep{sun2025fast,izzo2025ep250304a,srinivasaragavan2025ep250827b}. Additional events found with association to independently discovered supernovae also contribute, together giving a rate comparable to low luminosity GRBs \citep{liang2026archival}. Thus, we can conclude that, thus far, the overwhelming majority of Einstein Probe discovered transients are either gamma-ray bursts or jet-driven SBO events, and this adds to growing evidence that all Ic-BL arise from jetted engines.



\begin{figure}
    \includegraphics[width=0.5\textwidth]{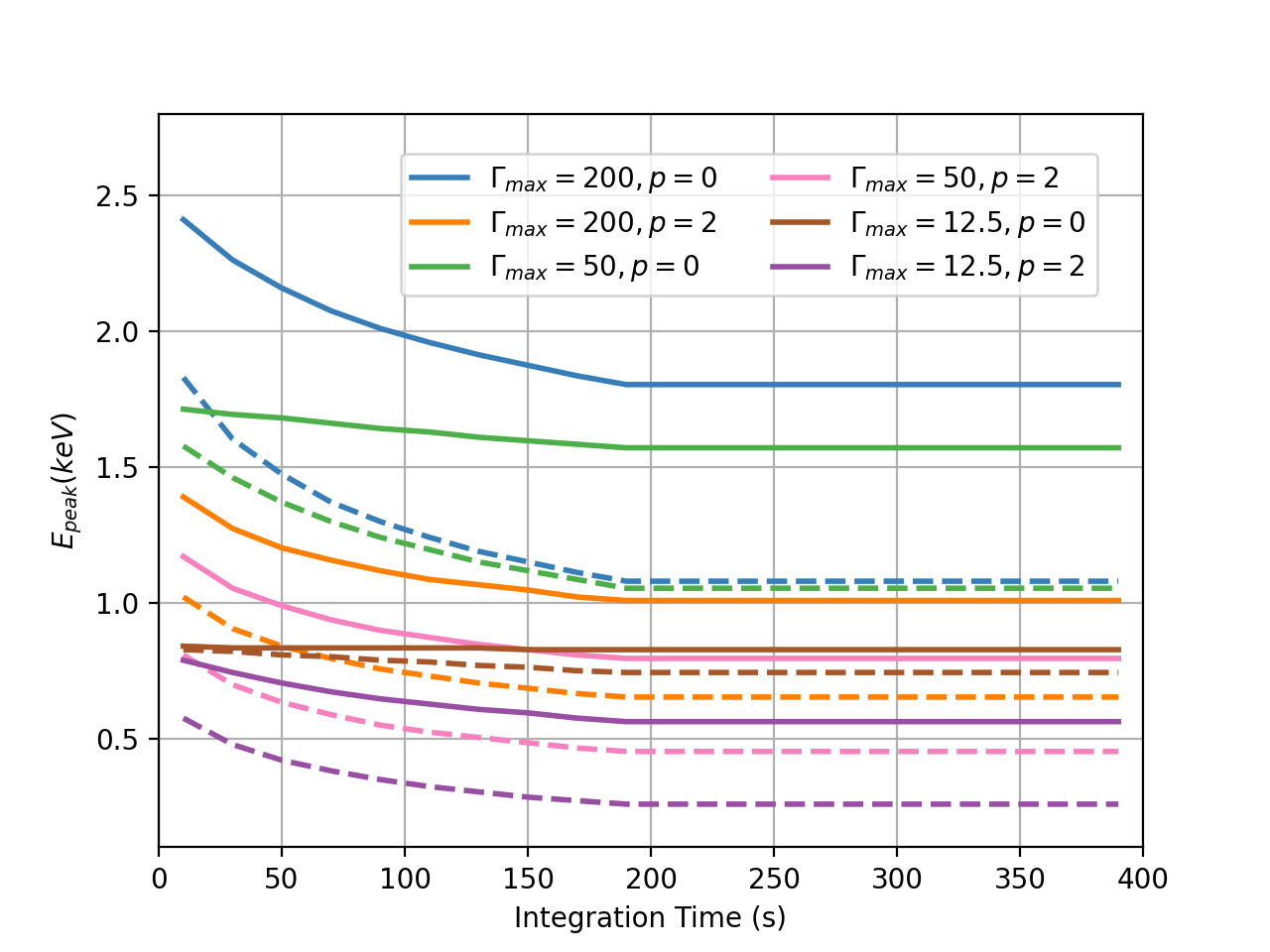}
    \caption{$E_{\rm peak}$ versus integrated time for a range of models varying the max Lorentz factor ($\Gamma_{\rm max}$), power-law index for the distribution of Lorentz factors ($p$), and total ejecta energy (solid $\equiv$ $10^{51}$\, erg, dashed $\equiv$ $10^{50}$\, erg).  In our simple model, $E_{\rm peak}$ is initially high and then decreases with time.  Although the observed $E_{\rm peak}$ depends on $\Gamma_{\rm max}$, it also depends on the total energy and the power-law index.  In addition, it also depends on the integration time of the observations.} 
    \label{fig:epeak}
\end{figure}

Many Einstein Probe papers focus on measuring or constraining the photon energy at peak emission ($E_{\rm peak}$).  $E_{\rm peak}$ can probe the maximum Lorentz factor.  But even with our simple models, there is not a direct connection between $E_{\rm peak}$ and $\Gamma_{\rm max}$.  Figure~\ref{fig:epeak} shows $E_{\rm peak}$ as a function of integration time varying Lorentz factor ($\Gamma_{\rm max}$), power-law index for the distribution of Lorentz factors ($p$), and total ejecta energy.  $E_{\rm peak}$ depends on all of these quantities, making it difficult to determine $\Gamma_{\rm max}$ directly.  Further, with our simple model, $E_{\rm peak}$ decreases with time.  Depending upon the duration of the observation, the value observed value of $E_{\rm peak}$ will be different for the same set of parameters.  Figure~\ref{fig:epeak} shows $E_{\rm peak}$ assuming the transient is observed immediately.  If there is a delay in the observation, the observed $E_{\rm peak}$ will be lower.  

\section{Projections for Current and Future Missions}
\label{sec:projections}
We here provide estimates for the detection rate of SBOs by current and future instruments. For each SBO type we provide a detection range and corresponding detection rate per instrument based on a pessimistic and an optimistic model. Thus, even measurement of the rate is informative. The instruments, their relevant properties, and their maximal detection distance ranges are given in Table~\ref{tab:instruments}. A summary of the key assumed intrinsic rates is given in Table~\ref{tab:volumetric-rates}. The combination of these with a presumed mission lifetime gives anticipated detection rates listed, summarized in Table~\ref{tab:detection-rates}. 

In order to quantify maximal detection distances and rates we utilize two different methods. For BlackCAT and derived instruments (MoonCAT, Mod 3, Mod 4) we emulate the full on-board trigger algorithm, detailed in Section~\ref{sec:blackcat}. For all other cases we utilize a basic comparison calculation of sensitive spacetime volume, accounting for source evolution, as detailed in Section~\ref{sec:AXIS}. We note that this refers only to detection, and makes no statement on the likelihood of classification and characterization, which generally require substantially more photon counts. Thus, the higher expected rates than historic identification rates is not surprising, and motivates real-time identification and follow-up to find the associated supernova.


To capture the extremes of our type Ia models, we use our ($A=10^{19} {\rm cm^2}$, $\rho= 10^{-6} \, {\rm g \, cm^{-3}}$) corresponding to our brightest model with a peak X-ray ($>$0.3\,keV) luminosity of $10^{43} {\rm \, erg \, s^{-1}}$ and peak photon energies in excess of 30\,keV.  The lower limit is given by the ($A=10^{20} {\rm cm^2}$, $\rho= 10^{-2} \, {\rm g \, cm^{-3}}$) model corresponding to a model with a peak X-ray ($>$0.3\,keV) luminosity of below $10^{40} {\rm \, erg \, s^{-1}}$ and peak photon energies around 10\,keV (see Figure~\ref{fig:flxray_Ia}).  For our type II bounding light-curves, we use the high energy model ($10^{48} {\rm \, erg}$ with $\Gamma \beta>0.1$) with peak Lorentz factor of 1.2 and a power law slope $p=6$ as an upper limit and the lower energy model ($10^{46} {\rm \, erg}$ with $\Gamma \beta>0.1$) with peak Lorentz factor of 1.1  and a power law slope $p=6$ as a lower limit.  These correspond to peak X-ray fluxes above 0.1\,keV of ($10^{43}, 10^{42} {\rm \, erg \, s^{-1}}$) respectively.  The peak emission for both of these models is around 0.1\,keV making them difficult to detect (Figure~\ref{fig:xrayII}).  For type Ib/c SNe, we assume the shock acceleration is much higher with peak Lorentz factors of 10 and power law slopes $p=4$.  The upper bound assumes $10^{50} {\rm \, erg}$ with $\Gamma \beta>0.1$ and the lower bound assumes $10^{48} {\rm \, erg}$ with $\Gamma \beta>0.1$.  These correspond to peak X-ray luminosities above 1.5\,keV of $3 \times 10^{44}, 10^{43} {\rm \, erg \, s^{-1}}$ respectively (Fig.~\ref{fig:flxray_area}).  Finally, for our jet models, we consider two bounding models with $10^{50} {\rm erg}$ energies with $\Gamma \beta > 0.1$ and flat ($p=0$) profiles.  The brighter model has a peak Lorentz factor of 500 and the dimmer model has a peak Lorentz factor of 100.  Both emit photons in excess of 1\,MeV.  Both models have peak luminosities with photon energies above 100\,keV above $10^{50} {\rm erg}$.

Many of these detections will, at best, be able to produce total emission in a few bands.  The final two columns of Table~\ref{tab:volumetric-rates} show ratios of emitted energy: $E_{1-10 {\rm \, keV}}/E_{0.1-1 {\rm \, keV}}$, $E_{10-1000 {\rm \, keV}}/E_{1-10 {\rm \, keV}}$ integrating the luminosity over 10\,s.  Not surprisingly, our type II models primarily emit in the UV and low-energy X-rays.  This explains why these systems are so difficult to detect in the X-ray.  A surprising result is that our Type Ib/c-BL ratios (supernovae produce from weak jets, $\Gamma<50$) have higher values than our jet models (seemingly more energy at high energies).  But this is an artifact of the fact that a)our jet models can peak above 1\,MeV and, more importantly, b) our jet models cool very quickly.  Over the course of 10s, the hottest ejecta from our jet models will lose much of their energy, emitting primarily at lower photon energies.  Our normal Type Ib/c supernovae with peak Lorentz factors below 50 take longer to cool and are still emitting strongly in the high-energy X-rays at 10s.  These results also stress the importance in SBO observations in devising instruments that can detect low-energy X-rays because, except in the case of compact stars, much of the emission occurs at these lower energies.

\begin{table*}
\begin{center}
\begin{tabular}{|c|c|c|c||c|c|c|c|c|c|}
\hline
Mission/ & Energy & Sensitivity & Field of  & II & Ia & Ibc & Ic-BL & Off-Axis  & On-Axis  \\ 
Instrument & Range [keV] & [erg/s/cm$^2$; s]& View [deg$^2$] & [Mpc] & [Mpc] & [Mpc] & [Mpc] & GRB [Mpc] & GRB [Mpc] \\ 
\hline
Swift-XRT   & 0.2-10 & 8E-14; 10,000 & 0.15 & 0.5-45 & 3-11 & 200-1000 & 1600-1900 & $>$7,800 & 5400-6100 \\
 \hline
EP-WXT  & 0.5-4 & 1E-10; 100 & 3600 & $<$0.4 & 0.7-2.2 & 20-150 & 350-460 & 2100-4500 & 690-1050 \\
Mod 1  & 0.5-4 & 8E-12; 100 & 300 & $<$1.5 & 2.6-7.5 &75-500 & 1200-1400 & $>$6000 & 3000-4300 \\
Mod 2  & 0.1-4 & 8E-12; 100 & 300 & 4.9-110 & 2.6-7.6 & 260-1250 & 1400-1900 & $>$8000 & 3100-4500 \\
 \hline
BlackCAT  & 1-20 & 8E-8; 10 & 2800 & 0 & 0.2-0.6 & 1.4-9.9 & 33 & 400-1000 & 300-330 \\
Mod 3  & 0.5-20  & 8e-9; 10 & 20,600 & $<$0.1 & 0.5-1.9 & 4-30 & 100-115 & 1400-3000 & 980-1080 \\
Mod 4  & 0.1-20  & 8e-9; 10 & 20,600 & 0.3-6.4 & 0.5-1.9 & 16-76 & 100-130 & 1500-3400 & 980-1090 \\
\hline
AXIS  & 0.5-2 / 0.3-10 & 2E-18; 7 Ms & 0.13 & $<$10 & 10-25 & 400-2000 & 5700-6600 & $>$10,000 & $>$10,000 \\
\hline
\end{tabular}
\caption{The active, forthcoming, and proposed instruments capable of detecting X-ray shock breakout, with their key capabilities and maximal detection distances for a range of source types. For each instrument we assume a 95\% livetime, except for EP-WXT where we use 50\% and BlackCAT with 76\%. Most instruments is considered to be signal dominated (linear flux sensitivity increase in time) except BlackCAT and derived instruments are background dominated (square root flux sensitivity increase in time). For AXIS the full range is 0.3-10 keV, while the 0.5-2 keV range is used for our quoted sensitivity. Modifications are described in Section~\ref{sec:mods}. The maximal detection distances are a range corresponding to the value for the dim and bright model in each case, where we note these numbers are considering only shock breakout emission (e.g., non-thermal emission in GRBs is neglected). Extinction is also neglected}
\label{tab:instruments}
\end{center}
\end{table*}

\begin{table*}
\begin{center}
\begin{tabular}{|c|c|c||c|c|}
\hline
Source Class & Volumetric Rate & Reference & $E_{1-10 {\rm \, keV}}/E_{0.1-1 {\rm \, keV}}$ & $E_{10-1000 {\rm \, keV}}/E_{1-10 {\rm \, keV}}$  \\
 & [Gpc$^{-3}$ yr$^{-1}$] & & & \\
\hline
Type II & 50,000 & \citet{li2011nearby} & $5\times 10^{-18}-7\times10^{-9}$ & $0.0-0.0$ \\
Type Ia & 30,000 & \citet{li2011nearby}  & $0.1-6$ & $10^{-5}-0.01$ \\
Type Ibc & 25,000 & \citet{li2011nearby}  & $0.002-0.006$ & $0.0005-0.0007$ \\
Type Ic-BL & 1,000 & \citet{corsi2023search} & $7-15$ & $0.1-6$ \\
Off-Axis GRB & 100 & \citet{ho2022landscape} & $0.5-0.8$ & $0.0001-0.003$ \\
On-Axis GRB & 1 & \citet{ho2022landscape} & $4-8$ & $0.2-0.7$ \\
\hline
\end{tabular}
\caption{The source classes considered with their corresponding volumetric rates. There is not yet full convergence in supernova rates due to the difficulty in determining efficiency functions and the expense of high-resolution spectroscopy for unambiguous classification. We take rough rates of from the quoted references, but note we vary some values for ease of future rescaling. The fraction of Ibc which are Ib and Ic is imprecise but of order 50\% \citep[e.g.][]{2017PASP..129e4201S}. While we do not make this separation here, Ic should generally be expected to have higher peak energies and luminosities.  The final 2 columns show two different emitted energy ratios assuming the SBO event is caught at the onset and integrated over 10s.}
\label{tab:volumetric-rates}
\end{center}
\end{table*}

\begin{table*}
\begin{center}
\begin{tabular}{|c|c|c|c|c|c|c|c|}
\hline
 & Mission Duration & Type II & Type Ia & Type Ibc & Type Ic-BL & Off-Axis GRB & On-Axis GRB \\
\hline
Swift-XRT & 21 years  & 0 & 0 & $<$0.4 & 0.01-0.01 & 2-5 & 0 \\
\hline
EP-WXT & 5 year  & 0 & 0 & 0.2-7.9 & 3.7-9.2 & 12,000-66,000 & 0.3-1.2 \\
Mod 1 & 3 years  & 0 & $<$0.001 & 1-250 & 21-33 & 5,000-35,000 & 4.7-13 \\
Mod 2 & 3 years  & $<$5.6 & $<$0.001 & 5.1-560 & 33-85 & 15,000-35,000 & 4.9-15 \\
\hline
BlackCAT & 1 year & 0 & 0 & 0–0.09 & 0.001 & 2–18 & 0.03–0.05 \\
Mod 3 & 3 years  & 0 & $<$0.001 & 0.01-4.4 & 0.6-0.9 & 19,000-300,000 & 6.8-9.0 \\
Mod 4 & 3 years  & $<$0.08 & $<$0.001 & 0.5-65 & 0.6-1.4 & 28,000-500,000 & 6.8-9.4 \\
\hline
AXIS & 5 years  & 0 & 0 & 0.2-25 & 3-5 & $\sim$24 & $\sim$0.24 \\
\hline
\end{tabular}
\caption{The expected detections of shock breakout from different progenitors by various instruments, again noting this considers only the thermal signature (i.e., non-thermal emission in GRBs is neglected). The range of values provided span the expectations for the faint model and the bright model, when available. The Ibc models include expectations for an SN~2008D model in parenthesis. The instruments and modified versions match those described in Table~\ref{tab:instruments}. While extinction is expected to strongly affect detections of Ia and II supernovae, the rates are so low this is unimportant. Particularly soft Ic-BL rates may be slightly lowered due to extinction.}
\label{tab:detection-rates}
\end{center}
\end{table*}

\subsection{AXIS}\label{sec:AXIS}
We evaluated the detectability of X-ray SBO transients from different progenitor types using the expected time-dependent sensitivity of the Advanced X-ray Imaging Satellite \citep[AXIS;][]{2023SPIE12678E..1ER}. For each event class—Type Ia, II, and Ibc supernovae (both bright and dim realizations), and GRBs--- we used model outputs providing the bolometric luminosity as a function of time and the time-resolved spectral energy distributions.

The luminosity curves were read and converted to physical units (erg/s). For each transient type, we derived the corresponding AXIS sensitivity curve as a function of integration time by scaling the nominal AXIS deep (7 Ms) field sensitivity in the 0.3--3\,keV band ($\sim 10^{-18}\,\mathrm{erg\,\,cm^{-2}\,s^{-1}}$, \citealt{2023SPIE12678E..1ER}) to shorter exposures.
We defined a special time (1 s for Type Ia, 100 s for Type Ibc and Type II SNe) based on the SBO peak and AXIS sensitivity scaling as a function of time to evaluate whether the predicted flux at that epoch would exceed the instantaneous AXIS sensitivity. To account for cosmological redshifting of the spectra, we applied an appropriate k-correction when converting intrinsic luminosities to observed fluxes:

\begin{equation}
F(t, D) = \frac{L(t)}{4\pi D^{2} (1 + z)}  k_{\mathrm{corr}}(D),
\end{equation}



AXIS would regularly detect a small population of Ic-BL SBO events as well as a sizable fraction of off-axis GRBs, being able to detect such signatures deep into the universe. AXIS may be able to detect a sample of Ibc supernova from convective engines, or provide a non-detection sufficient to constrain most viable models. For AXIS to see type II or Ia supernovae, it would require rapid response to early warning reports from neutrino or gravitational wave detectors.


\subsection{Swift-XRT}
We repeat a similar produce for Swift-XRT. The chance of detecting one Ibc in 21~years is about 1\%, providing SN~2008D as a lucky event. We anticipate a few off-axis GRB detections which may have escaped identification due to high latency for the searches for transients in XRT \citep{evans20202sxps} as well as lack of confirmation due to detection deep into the universe.

\subsection{Einstein Probe}
As previously discussed, Einstein Probe (EP) is an X-ray telescope designed to identify fast X-ray transients \citep{yuan2022einstein}. The Wide-field X-ray Telescope (WXT) is the most prolific detector of fast x-ray transients launched, with a rate of 70-80 events announced over GCN each year \citep{o2025redshift}, which are not expected to arise from stellar flares (the greatest source of contamination). The majority of events with substantial follow-up are associated to Ic-BL supernovae and/or gamma-ray bursts. We estimate Einstein Probe detections through the method described above.

The Einstein Probe team gives a rate of the Ic-BL SBO events of $\sim$30-76~Gpc$^{-3}$yr$^{-1}$ \citep{li2025extremely,liang2026archival}. This is comparable to the rate of low luminosity GRBs, further suggesting a connection. The most distant of these events, EP240414a, is at about 2~Gpc, consistent with our off-axis detection distance. Our annual rates appear to be in excess of the identification rate of these transients. However, this is to be expected due to the losses due to follow-up efficiency, as well as the majority of Einstein Probe detections being subthreshold. The joint X-ray and optical recovery of EP250827b \citep{srinivasaragavan2025ep250827b} and EP240506a \citep{liang2026archival} demonstrates joint searches can classify additional events, but this success rate fails precipitously at cosmological distances. If Einstein Probe completes its 5 year mission with no identification of a Ibc (non-broad lined) supernova, this will be informative on the SBO events of convective engine supernova.


\subsection{BlackCAT}\label{sec:blackcat}

BlackCAT is a NASA CubeSat mission designed to monitor the X-ray sky for GRBs and other high-energy transients launched in January 2026 ~\citep{falcone2024blackcat}.
This mission uses an X-ray coded-aperture telescope to provide a large field of view (0.85~sr partially coded) and sub-arcminute localization capabilities.

We simulate BlackCAT observations of various X-ray SBO transients using techniques similar to those described in \citet{2024ApJ...969..138C}.
Type~Ia and Type~II SNe events are sampled from a uniform-density distribution to a maximum distance of 30 Mpc. 
Type~Ibc~SNe are sampled from a uniform-density distribution extending to 100 Mpc. 
Low-luminosity GRBs are drawn from the \citet{madau2014cosmic} star formation history model with a maximum redshift of 0.4.

In the simulation, sources are randomly placed within the instrument field of view.
Photons are sampled from the source and from cosmic and galactic backgrounds, then processed through the instrument's response.
These simulated photons are fed into an imaging trigger algorithm, separated into 7 time bins between 0.125 and 64~s and 10 energy bins between 1.0 and 20.0 keV.
An event is considered a detection if any one of the trigger bins produces a peak at the expected position in the recreated sky image exceeding a set threshold.
Due to the high trials factor of the instrument, we use a threshold for independent detection of $6.5\sigma$, relative to the nearby background.

\subsection{Other Wide-field Monitors}\label{sec:mods}
We create a limited set of modified instrument designs in order to understand the limits of current technology, were missions designed specifically for SBO. AXIS is already an extreme version of maximal sensitivity, though with a smaller field of view than typical transient monitors. 

With modular designs one can trade of larger field of view for copointed detectors, trading coverage for depth. When the depth has not reached the sensitivity to cover the peak rate of these events (e.g. redshift 1-2 for massive star transients), depth results in a higher detection rate. We thus consider a version of the Einstein Probe WXT where the 12 lobster optics are coaligned, referred to as Mod 1. An additional exploration is whether a lower low energy threshold would be capable of recovering type Ia or type II SBO (neglecting extinction). Mod 2 refers to a coaligned WXT with a low energy threshold of 0.1~keV. Mod 3 refers to an extension of BlackCAT that is ten times more sensitive as well as having a larger field of view and livetime, and Mod 4 to the same but with the 0.1~keV low energy threshold.

We conclude that many monitors can recover jetted explosion SBO events. With current technologies tasked with SBO as a priority it may be possible to recover convective engine stripped envelope SBO, giving unique insights into massive stars at the end of their lives, though detections may still remain rare if these signals are on the faint end of our predictions. Lastly, we would require great fortune to serendipitously detect type Ia or type II without large improvements in detector capability.

\subsection{Early Warning Response}
Each of these transient classes can, in principle, be first detected by gravitational waves or neutrinos. The forthcoming megaton neutrino detectors, JUNE \citep{abusleme2021juno}, Hyper-K \citep{abe2018hyper}, and DUNE \citep{falcone2022deep} can detect neutrinos from the core-collapse of massive stars which result in neutron stars, and comparable neutrino signals from accretion in jet-driven events, in the Milky Way and out to Andromeda. Similarly, third generation gravitational wave interferometers, being Einstein Telescope \citep{branchesi2023science} and Cosmic Explorer \citep{evans2021horizon}, can detect convective engine events in the Milky Way and jetted supernova to perhaps a few tens of Mpc. 

While these distances are close, resulting in rates lower than one per decade, SBO detection would provide unique information on the explosion and the stellar progenitor, and is the first electromagnetic signal and thus the first opportunity for a precise location for broadband follow-up of transformational events. The breakout time can be anywhere from a minute to hours. Any X-ray telescope would ideally be pointed at any gravitational wave or neutrino detection of such an event; however, the localization area will generally exceed the field of view of narrow-field X-ray telescopes. This may be overcome with host galaxy or supergiant catalogs in certain situations.

Lastly, we comment on early warning for type Ia. It is generally believed that most of these events arise from the merging of two white dwarfs, making them detectable to mid-range gravitational wave interferometers. Mission concepts would be capable of regular detections of WD-WD mergers, thought to be the dominant progenitor channel for type Ia supernovae, and giving localizations precise and early enough for narrow-field X-ray telescopes to provide coverage \citep[e.g.][]{kinugawa2022probe}. A telescope like AXIS would be ideal for multimessenger detections with SBO.

\section{Summary}
\label{sec:summary}

In this paper, we introduced a semi-analytic code that calculates the Brehmsstrahlung emission from shock breakout for a broad range of astrophysical transients.  This code takes an assumed distribution of ejecta velocities and calculates the Brehmsstrahlung emission as a function of time.  This tool is open-sourced and can be found at \url{https://github.com/clfryer/SBO}.  This code includes special relativistic effects on both the shock properties and relativistic beaming.

With this tool, we calculated signals for a range of SBO events from thermonuclear and core-collapse supernovae to long-duration gamma-ray bursts and broad-line supernovae.  We present results showing both light-curves in different bands and spectral evolution with time.  Due to uncertainties in the initial conditions, a broad range of results are possible.  We validate these results against observed SBO and potential SBO events.  At this time, the current models generally fit existing observations with reasonable initial conditions within our model framework.  As the number of observed events increases, the distribution of events will guide the conditions in the explosions (both the progenitor and explosion properties).  Any event that can not be explained by our current framework will point to new insight into astrophysical transients.

Finally, with these models, we can make predictions for source detection rates for recent and upcoming observatories.   Unless our current understanding of type II supernovae is incorrect, it is unlikely that we will detect this class of supernovae in the X-ray.  Although not studied here, ultraviolet observations are likely to be the better diagnostic for SBO of type II supernovae.  Similarly, unless shock interactions in the complex binary progenitor of type Ia supernovae can drive strong X-ray emission, it is unlikely X-rays will detect SN Ia SBO.  But X-ray detectors should detect SBO from Ibc, Ic-BL and GRB transients.  It is important to stress that characterizing the emission (not just detecting these events, but observing spectral properties) will be critical in probing the explosions and progenitors of these transients.

Our future work will focus on ultraviolet signatures and studying the effects of shock interactions.

\begin{acknowledgements}

The work by CLF was supported by the US Department of Energy through the Los Alamos National Laboratory. Los Alamos National Laboratory is operated by Triad National Security, LLC, for the National Nuclear Security Administration of U.S.\ Department of Energy (Contract No.\ 89233218CNA000001). JC's research was supported by an appointment to the NASA Postdoctoral Program at the NASA Goddard Space Flight Center, administered by Oak Ridge Associated Universities under contract with NASA. BO gratefully acknowledges support from the McWilliams Fellowship at Carnegie Mellon University.

\end{acknowledgements}

\bibliography{refs}{}
\bibliographystyle{aasjournal}

\end{document}